\documentclass[12pt,preprint]{aastex}
\usepackage{lscape}
\begin{document}

\title{Revealing the Broad Line Region of NGC 1275: The Relationship to Jet Power}
\author{Brian Punsly\altaffilmark{1}, Paola Marziani\altaffilmark{2}, Vardha N. Bennert\altaffilmark{3},
Hiroshi Nagai\altaffilmark{4,5} and Mark A.
Gurwell\altaffilmark{6}}\altaffiltext{1}{1415 Granvia Altamira,
Palos Verdes Estates CA, USA 90274: ICRANet, Piazza della Repubblica
10 Pescara 65100, Italy and ICRA, Physics Department, University La
Sapienza, Roma, Italy, brian.punsly@cox.net}\altaffiltext{2}{INAF,
Osservatorio Astronomico di Padova, Italia}\altaffiltext{3}{Physics
Department, California Polytechnic  State  University, San Luis
Obispo, CA 93407, USA}\altaffiltext{4}{National Astronomical
Observatory of Japan, Osawa 2-21-1, Mitaka, Tokyo 181-8588,
Japan}\altaffiltext{5}{The Graduate University for Advanced Studies
(SOUKENDAI), Osawa 2-21-1, Mitaka, Tokyo 181-8588,
Japan}\altaffiltext{6}{Harvard-Smithsonian Center for Astrophysics,
Cambridge, MA USA}

\begin{abstract}
NGC 1275 is one of the most conspicuous active galactic nuclei (AGN)
in the local Universe. The radio jet currently emits a flux density
of $\sim 10$ Jy at $\sim 1$ mm wavelengths, down from the historic
high of $\sim 65$ Jy in 1980. Yet, the nature of the AGN in NGC 1275
is still controversial. It has been debated whether this is a broad
emission line (BEL) Seyfert galaxy, an obscured Seyfert galaxy, a
narrow line radio galaxy or a BL-Lac object. We clearly demonstrate
a persistent H$\beta$ BEL over the last 35 years with a full width
half maximum (FWHM) of 4150 - 6000 km/s. We also find a prominent
P$\alpha$ BEL (FWHM $\approx 4770 $ km/s) and a weak CIV BEL (FWHM
$\approx 4000 $ km/s), H$\beta$/CIV $\approx 2$. A far UV HST
observation during suppressed jet activity reveals a low luminosity
continuum. The H$\beta$ BEL luminosity is typical of broad line
Seyfert galaxies with similar far UV luminosity. X-ray observations
indicate a softer ionizing continuum than expected for a broad line
Seyfert galaxy with similar far UV luminosity. This is opposite of
the expectation of advection dominated accretion. The AGN continuum
appears to be thermal emission from a low luminosity, optically
thick, accretion flow with a low Eddington ratio, $\sim 0.0001$. The
soft, weak ionizing continuum is consistent with the relatively weak
CIV BEL. Evidence that the BEL luminosity is correlated with the jet
mm wave luminosity is presented. Apparently, the accretion rate
regulates jet power.
\end{abstract}


\section{Introduction} The galaxy NGC 1275 was distinguished from
other extragalactic objects as an active galactic nucleus (AGN) due
to its bright stellar like nucleus and strong emission lines
\citep{hum32,sey43}. It lies at the heart of the Perseus cluster of
galaxies. It has been the brightest extragalactic radio source (3C
84) at high frequency with flux densities of 45Jy - 65Jy in 1980 at
270 GHz and $\sim 40-50$ Jy at 90 GHz from 1965-1985 \citep{nes95}.
The large radio flare began in the 1950s and was accompanied by
large optical flares in the 1970s \citep{nes95}. The flare subsided
in the 1990s, but a new strong flare began in 2005. The physical
nature of the AGN is of great interest due to the powerful radio jet
and its relationship to the feedback in the cluster cooling flow.
\par In spite of its proximity ($z = 0.0175$) and unique properties,
the nature of the AGN has been the subject of controversy within
standard classification schemes that are designed to segregate
physical characteristics. In the original classification scheme, it
was considered a Seyfert 2 galaxy \citep{kha74,chu85}. Noting the
wildly varying optical flux, it was later called a BL-Lac object
\citep{ver78}. Reports of a weak broad emission line (BEL) region
based on H$\alpha$ were later made with better spectra and a
classification of Seyfert 1.5 was assigned \citep{fil85,lho97}.
However, BELs were not detected by other observers with the same
telescope, but they did conclude that it could not be ruled out due
to insufficient signal to noise \citep{ros94}. More recently, the
general notion of identifying emission lines as weak BELs as opposed
to wide narrow lines (NLs) in AGN has been the subject of
controversy. To resolve the issue, narrow (0.1"-0.2") slit HST
observations of many of the radio galaxies in \citet{lho97} were
performed. The small aperture excludes much of the H$\alpha$ NL
emission. This showed that many of these radio galaxies did not have
a BEL in H$\alpha$ \citep{bal14}. Thus, NGC 1275 is often called a
NL radio galaxy. Even so, its classification as a low or high
excitation radio galaxy (i.e., does it have a strong
photo-ionization source in the accretion flow?) is controversial
\citep{but10,son12}. From a physical point of view, these
distinctions are critical. Is the powerful radio jet ejected from a
hot advection dominated accretion flow (ADAF), \citet{nar94}, or
does it emanate from an accretion flow that cools by thermal
emission from optically thick gas (that photo-ionizes the
surrounding gas)?

\begin{figure}
\begin{center}
\includegraphics[width=135 mm, angle= 0]{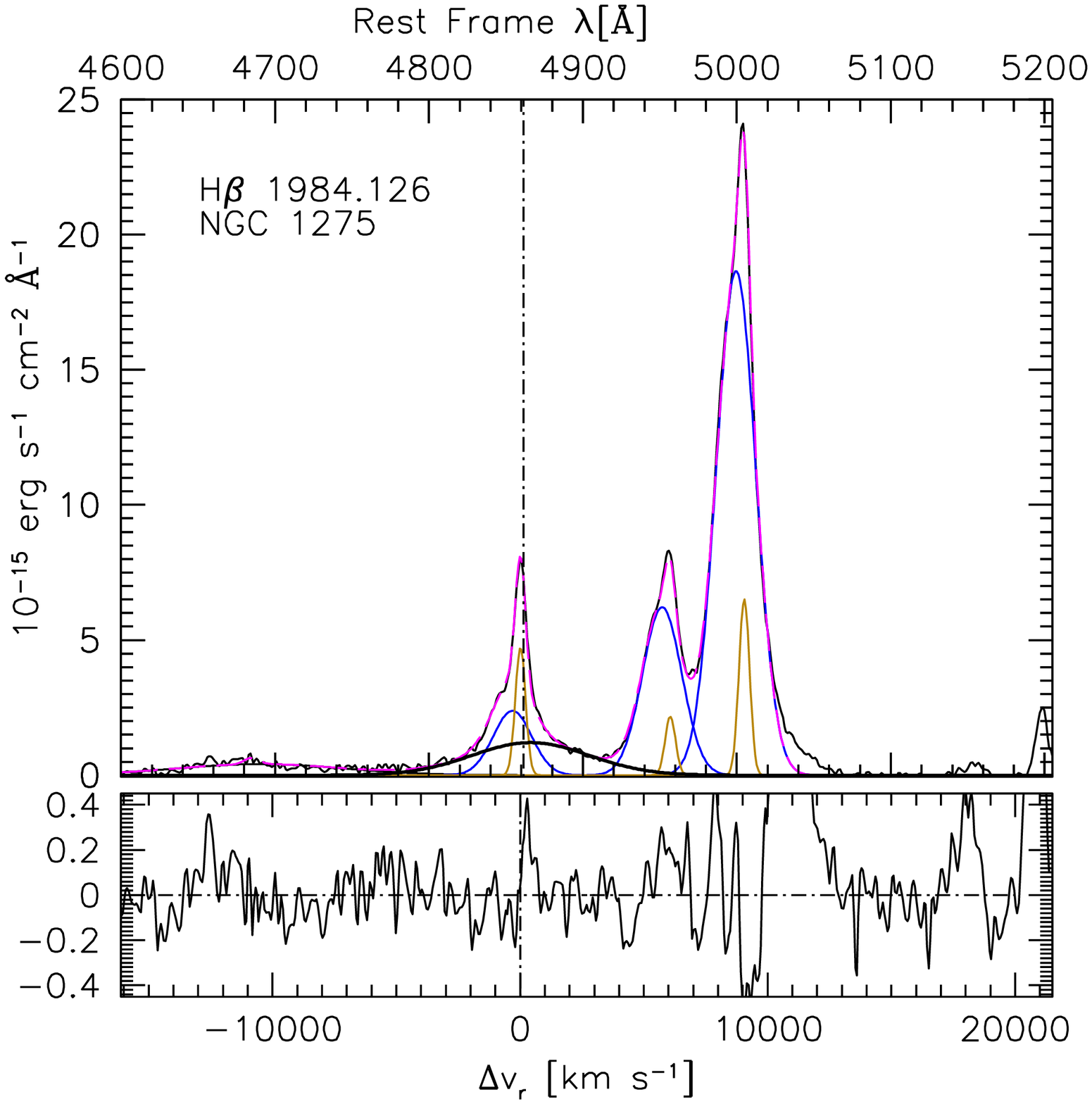}
\includegraphics[width=65 mm, angle= 0]{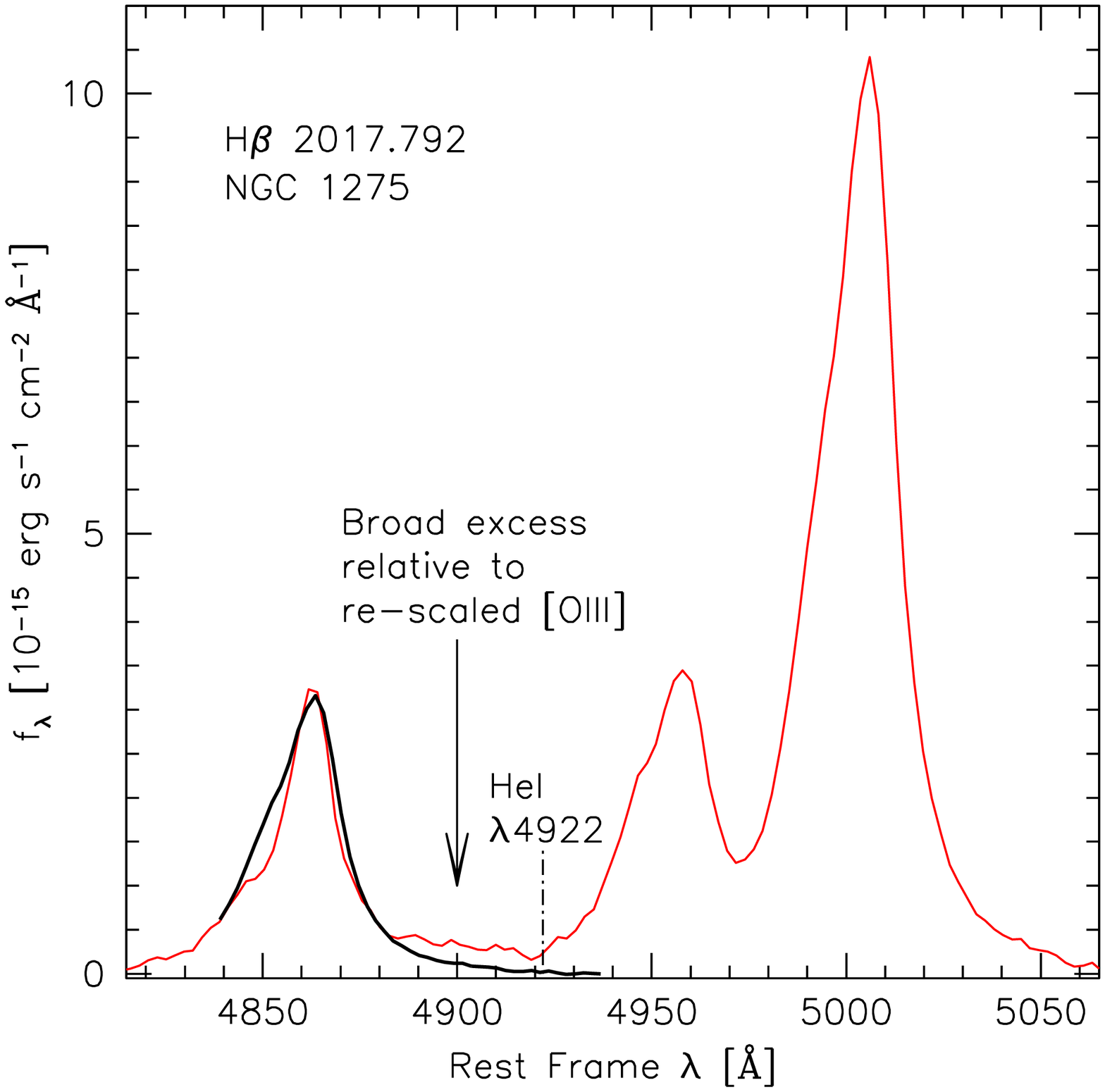}
\includegraphics[width=65 mm, angle= 0]{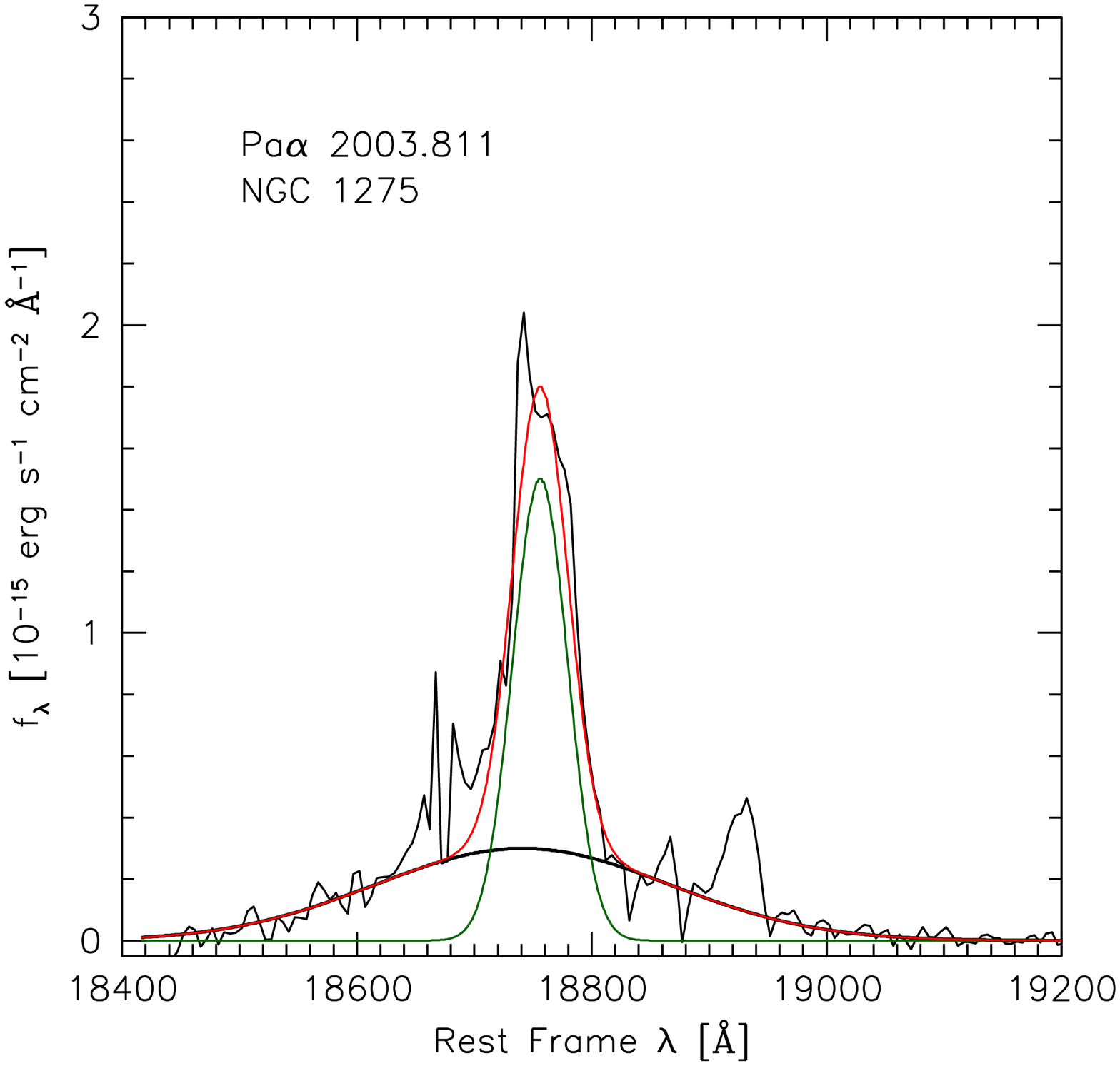}
\caption{The top frame illustrates our fitting procedure for the
H$\beta$ and [OIII] complex (epoch 1984.126) that is described in
the text. The BEL, semi-broad NL, and NL (wind) are black, blue and
yellow, respectively. The bottom left frame shows the redwing excess
of H$\beta$ relative to a scaled [OIII] NL profile. Note that there
is no evidence of a weak HeI $\lambda$4922 emission line or the FeII
$\lambda$4924 emission line. The bottom right frame shows the BEL
(NNL) in P$\alpha$ in brown (green).}
\end{center}
\end{figure}
\par Thus motivated, we look for evidence of the BEL in high signal
to noise spectra both historical and with new spectral observations.
We concentrate on $H\beta$ rather than $H\alpha$ even though it is
considerably weaker and potentially suffers from FeII contamination.
These issues are less severe than subtracting [NII] from the
$H\alpha$ + [NII] complex. Subtracting [NII] from the composite
profile using other strong NL profiles is nontrivial and is highly
dependent on the NL profile ([SII] or [OI]) that is utilized to
scale the subtraction \citep{bal14}. In Section 2 and 3, we describe
the BELs. Section 4 is a discussion of the ionizing continuum and
the Section 5 describes the BEL - jet connection.

\section{Simultaneous Fits of H$\beta$ and [OIII]$\lambda\lambda 4959,\, 5007$}
In this analysis, it is very important to fit self-consistently
H$\beta$ and [OIII]$\lambda\lambda 4959,\, 5007$. The models were
carried out assuming two NL components: a semi broad NL and a
narrower NL (perhaps of galactic wind origin or unperturbed bulge
gas) for both [OIII] and H$\beta$. For the [OIII] doublet, we adopt
the usual condition that the 4959 $\AA$ line flux is 1/3 of the 5007
$\AA$ line flux and the two components have the same velocity and
FWHM \citep{dim07}. The same velocity and FWHM were applied to the
corresponding two NL components of H$\beta$. The FeII subtraction is
accomplished using an FeII template based on I Zw 1, see
\citet{mar09}, Section 2.3. The FeII$\lambda 4570$ blend is
undetectable \citep{bor92}. Thus, the template indicates that
FeII$\lambda 4924$ is very weak and won't affect our fit (see Figure
1). For the sake of scientific rigor all the data need to be fit
with the same methodology. Previously published results are refit.

\par Table 1 describes the line fits with complete expository text and
labels. All of the data is archival except for our new observation
in 2017. Long slit spectra were obtained with the AFOSC camera
mounted at the Cassegrain focus of the Copernico telescope on Mount
Ekar, on Oct. 18, 2017 and Oct. 26, 2017 (Grism \# 7). The slit was
opened to 1.69", and  2.50" at parallactic PA and at PA=90, and
spectra were extracted  as reported in Table 1.  The GR \# 7 yields
an inverse resolution of $\approx 590$ for a slit width 1.69'',
allowing for an effective deblending of the BC from the wind and
narrow components. The 2D frames of the spectra were bias
subtracted, flat fielded, wavelength and flux calibrated. The
wavelength calibration obtained  from lamps of Th and Ar was
realigned on the sky lines of the source frame, yielding a
dispersion in residual sky line shifts with respect to rest
wavelength of $0 \pm 25$ km/s. The flux calibration was achieved by
the observations of tertiary standard stars obtained with the same
slit width right before and after NGC 1275. The adopted absolute
flux scale at fixed aperture is believed to be accurate to about
$10$\% at a $1\sigma$ confidence level.

\par The top frame of Figure 1 shows the fit to the strongest H$\beta$
BEL. The BEL is not obvious because it does not dominate the NL
contribution to the combined profile. The bottom two frames in
Figure 1 corroborate the BEL identification. For illustrative
purposes, the bottom left frame shows the residual excess in the
redwing that occurs if one tries to fit H$\beta$ with NLs only. If
we simply re-scale the amplitude of [OIII]$\lambda 5007$ and then
center it on H$\beta$, the figure shows a `` red shelf" of excess
flux. Mathematically, we performed an F-test between the best fit
with and without a BEL. For 641 degrees of freedom, the F-statistic
was 2.768 which implies a better fit with the BEL at $>99.999\%$
significance level. The result derives primarily from residuals
below the ``red shelf". The bottom right hand frame shows the
P$\alpha$ line which also arises from transitions from the n=4
state. There is a clear BEL with a similar FWHM to the H$\beta$ BELs
in Table 1, 4770 km/s.
\par The line luminosity in Table 1 was computed using the following cosmological
parameters: $H_{0}$=71 km/s/Mpc, $\Omega_{\Lambda}=0.73$ and
$\Omega_{m}=0.27$ and $z=0.0175$. Notice that we fit [OII]$\lambda
3727$ as well as CIV and P$\alpha$. Line luminosity is corrected for
Galactic extinction. The best fit to the extinction values in the
NASA Extragalactic Database (NED) in terms of \citet{car89} models
is $A_{V}=0.438$ and $R_{V}=3.0$.
\begin{landscape}
\begin{table}
\caption{NGC 1275 Emission Line Fits}
{\tiny\begin{tabular}{cccccccccccc} \tableline\rule{0mm}{3mm}
  &    &  Optical &    &   & Emission &   &  & Lines &  & & \\
\tableline \rule{0mm}{3mm}
1 &  2  &  3 &  4  & 5  & 6 & 7 & 8 & 9 & 10 & 11&12 \\
\tableline \rule{0mm}{3mm}
 Date &  Extraction & $H\beta$ (broad) & $H\beta$ (broad) &  [OIII] (NL)\tablenotemark{a} & [OIII] (NL)\tablenotemark{a} & [OIII] galactic)\tablenotemark{a} & [OIII] (galactic)\tablenotemark{a}  & $H\beta$ (NL) & $H\beta$ (galactic)& [OII] & Reference\\
 & Region  &  Luminosity & FWHM & Luminosity & FWHM & Luminosity &FWHM & Luminosity & Luminosity & Luminosity &\\
 &  Arcsec  &  ergs/s &  km/s  & ergs/s  & km/s & ergs/s &km/s & ergs/s & ergs/s  & ergs/s & \\
\tableline \rule{0mm}{3mm}
1983.940\tablenotemark{b} & 2 x 128\tablenotemark{c} & $1.28\times 10^{41}$ & $5161\pm 33$ & $6.20\times 10^{41}$  & $1825\pm 5$ & $2.32\times 10^{40}$ & 482 & $6.27 \times 10^{40}$ &$3.96\times 10^{40}$ &$2.49\times 10^{41}$&\citet{law96}\\
1984.126\tablenotemark{b} & 2 x 4\tablenotemark{c} & $1.28\times 10^{41}$ & $5536\pm 196$ & $6.52\times 10^{41}$  & $1821\pm 6$ & $6.06\times 10^{40}$ & 484 & $8.32 \times 10^{40}$ &$4.33\times 10^{40}$ & ...& \citet{lho95}\\
1994.940\tablenotemark{e} & $\approx$2.6 x 3\tablenotemark{c} & $7.88\times 10^{40}$ & $5162\pm 275$ & $5.05\times 10^{41}$  & $1754\pm 7$ & $6.57\times 10^{40}$ & 588 & $5.70 \times 10^{40}$ &$4.46\times 10^{40}$ &$3.55\times 10^{41}$& \citet{mar96}\\
2000.663\tablenotemark{f} & 0.2 x 0.2 & $6.10\times 10^{40}$ & $4169\pm 299$ & $3.81\times 10^{41}$  & $1907\pm 17$ & $5.42\times 10^{38}$ & 897 & $4.99 \times 10^{40}$ &$1.44\times 10^{40}$ & $8.40\times 10^{40}$& This paper\\
2008.016\tablenotemark{g,h} & 2 x 2 \tablenotemark{d}& $9.21\times 10^{40}$ & $4260\pm 138$ & $3.62\times 10^{41}$  & $2046\pm 23$ & $5.73\times 10^{40}\tablenotemark{h}$ & 1115 & $2.66 \times 10^{40}$\tablenotemark{h} &$4.11\times 10^{40}$\tablenotemark{h}&$2.42\times 10^{41}$ & \citet{but09}\\
2009.646\tablenotemark{i} & 2 x 2 \tablenotemark{d}& $8.77\times 10^{40}$ & $4556\pm 75$ & $3.67\times 10^{41}$  & $1774\pm 5$ & $5.77\times 10^{40}$ & 560 & $4.45 \times 10^{40}$ &$3.42\times 10^{40}$&$2.00\times 10^{41}$ & \citet{son12}\\
2009.646\tablenotemark{i} & 2 x 3 \tablenotemark{d}& $9.66\times 10^{40}$ & $4475\pm 77$ & $4.32\times 10^{41}$  & $1768\pm 5$ & $7.03\times 10^{40}$ & 551 & $5.12 \times 10^{40}$ &$4.08\times 10^{40}$&$2.41\times 10^{41}$ & \citet{son12} \\
2009.646\tablenotemark{i} & 2 x 4 \tablenotemark{d}& $9.41\times 10^{40}$ & $4470\pm 87$ & $4.60\times 10^{41}$  & $1764\pm 5$ & $7.94\times 10^{40}$ & 542 & $5.84 \times 10^{40}$ &$4.37\times 10^{40}$&$2.66\times 10^{41}$ & \citet{son12} \\
2017.792\tablenotemark{j} & 1.69 x 2 \tablenotemark{d}& $8.18\times 10^{40}$ & $5973\pm 322$ & $2.92\times 10^{41}$  & $1779\pm 17$ & $4.76\times 10^{40}$ & 590 & $3.66 \times 10^{40}$ &$2.93\times 10^{40}$ & ... & This paper\\
2017.792\tablenotemark{j} & 2.5 x 2.5 \tablenotemark{d}& $7.81\times 10^{40}$ & $4667\pm 137$ & $3.92\times 10^{41}$  & $1809\pm 7$ & $3.93\times 10^{40}$ & 710 & $4.88 \times 10^{40}$ &$3.91\times 10^{40}$ & ... & This paper\\
2017.792\tablenotemark{j} & 2.5 x 2.5\tablenotemark{c}& $1.05\times 10^{41}$ & $4967\pm 106$ & $4.14\times 10^{41}$  & $1805\pm 7$ & $7.26\times 10^{40}$ & 690 & $4.88 \times 10^{40}$ &$4.52\times 10^{40}$ & ...& This paper\\
2017.792\tablenotemark{j} & 2.5 x 3.5\tablenotemark{c} & $9.28\times 10^{40}$ & $4395\pm 184$ & $4.65\times 10^{41}$  & $1786\pm 9$ & $8.45\times 10^{40}$ & 678 & $5.86 \times 10^{40}$ &$4.88\times 10^{40}$ & ... & This paper\\
\tableline \rule{0mm}{3mm}
  &    &  Other &    &   & Emission &   &  & Lines &  &  &\\
\tableline \rule{0mm}{3mm}
1 &  2  &  3 &  4  & 5  & 6 & 7 & 8 & 9 & 10 & 11 &12 \\
\tableline \rule{0mm}{3mm}
 Date &  Extraction & $P\alpha$ (broad) & $P\alpha$ (broad) & $P\alpha$ (NL) & $P\alpha$ (NL) & CIV (broad) & CIV (broad)  & CIV (NL) & CIV (NL)& HeII (broad) & Reference \\
 & Region  &  Luminosity & FWHM & Luminosity & FWHM & Luminosity &FWHM & Luminosity & FWHM & Luminosity &\\
 &  Arcsec  &  ergs/s &  km/s  & ergs/s  & km/s & ergs/s &km/s & ergs/s & km/s  & ergs/s & \\
 \tableline \rule{0mm}{3mm}
2003.811\tablenotemark{k} & 0.8" x 15" & $6.56\times 10^{40}$ & $4770\pm 200$ & $6.12\times 10^{41}$  & $880\pm 175$ & ... & ... & ... &... & ...& \citet{rif06} \\
1993.093\tablenotemark{l} & 0.25" x 1.23" & ... & ... & ...  & ... & $3.39\times 10^{40}$ & $4297\pm 974$ & $5.09 \times 10^{40}$ &$1629 \pm 40$ & ...& \citet{mar96} \\
2011.352\tablenotemark{m} & 2" x 2" & ... & ... & ...  & ... & $4.22\times 10^{40}$ & $4477\pm 933$ & $6.82 \times 10^{40}$ &$1431 \pm 70$ &$ 3.14\times 10^{40}$ & This paper \\
\end{tabular}}
\tablenotetext{a}{$\lambda=5007$, total luminosity of doublet is
found by multiplying by 1.33. NL is the narrow line associated with
the AGN and galactic is large scale galactic gas emission}
\tablenotetext{b}{Hale Telescope} \tablenotetext{c}{Slit along PA =
90 degrees} \tablenotetext{d}{Slit at parallactic
angle}\tablenotetext{e}{San Pedro Martir 2.2
Meter}\tablenotetext{f}{HST STIS G430L}\tablenotetext{g}{Telescopio
Nazionale Galileo, 3.58 meter}\tablenotetext{h}{low resolution,
$20\AA$: characterizations of NLs not accurate and FWHMs not
accurate especially for NLs}\tablenotetext{i}{Lick 3
meter}\tablenotetext{j}{Copernico Telescope, 1.8 Meter. The
observations were performed for the purposes of this
study}\tablenotetext{k}{NASA 3m Infrared Telescope Facility }
\tablenotetext{l}{HST FOS G130H} \tablenotetext{m}{HST COS G160M.
Guide star acquisition failed. Thus, CIV luminosity possibly
underestimated due to slit losses from poor centering late in the
observation.}
\end{table}
\end{landscape}
It is important to note that the observations in Table 1 were not
preferentially selected because they showed a BEL. These were all
the available data-sets that could be obtained from data archives or
through direct contact with principal investigators. We know of no
other data that is well calibrated with good signal to noise for
which the data was available. For all the observations in Table 1,
the addition of a broad component improved the fit because of the
issue of the excess illustrated by the lower left hand frame of
Figure 1 and the related discussion.

\par Table 1 contains two important experiments based on the 2009 and
2017 data. Different extraction regions were applied to the data and
in 2017 different slit orientations were tried. The 2017 experiment
indicates that the H$\beta$ BEL fits were not driven by slit angle
or extraction region as long as the extraction region was larger
than the dimensions set by ``seeing". Thus, the values in Table 1
are robust for comparing the time evolution of the H$\beta$ BEL.
Conversely, the NL luminosity scales with the size of the extraction
region.
\section{The CIV Broad Emission Line} The top frame of Figure 2
shows that a NL cannot fit the 2011 CIV emission line. There is a
significant excess in the red wing and a smaller excess in the blue
wing that cannot be removed by raising the continuum level. The
bottom frame shows fits to two epochs obtained by adding a BEL. The
BEL+NL model fits are not well constrained by the noisy data and the
data gap in 2011. We remove some arbitrariness by requiring that the
NL FWHM agree to within 15\% between the two epochs. The results are
shown in Table 1 and Figure 2. In spite of the aforementioned data
quality, it is apparent that the BEL is extremely weak, even
compared to the weak $H\beta$ BELs. Based on Table 2, the ratio of
BEL luminosity L(CIV)/L($H\beta$) $\approx$ 1/2. We compare this to
other ``Population B" AGN defined by a FWHM of $H\beta$ that exceeds
4000 km/sec, as is the case for NGC 1275 \citep{sul00}. For
Population B AGN, typically, L(CIV)/L($H\beta$) $\approx$ 5-7
\citep{sul07}.

\begin{figure}
\begin{center}
\includegraphics[width=100 mm, angle= 0]{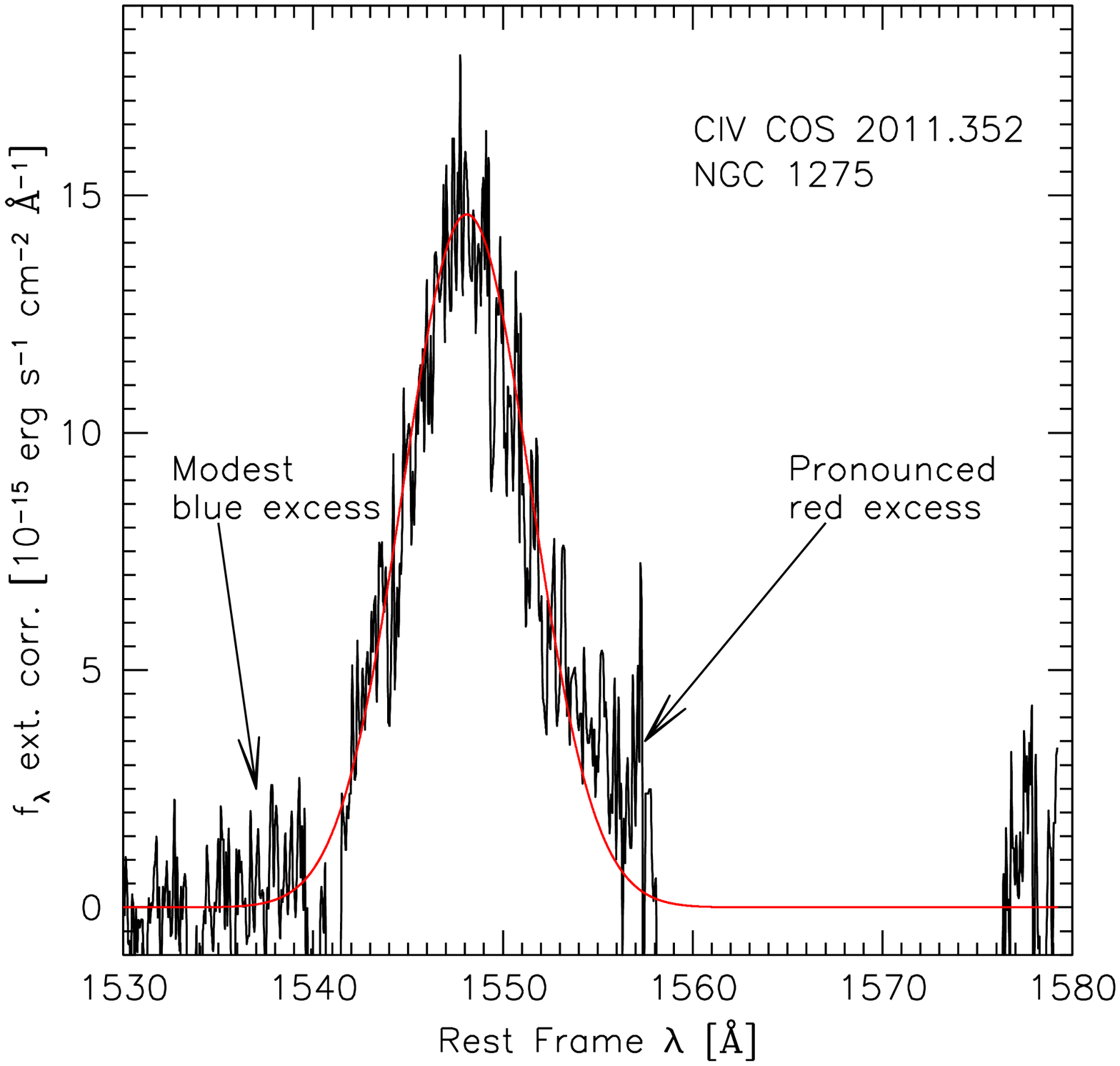}
\includegraphics[width=75 mm, angle= 0]{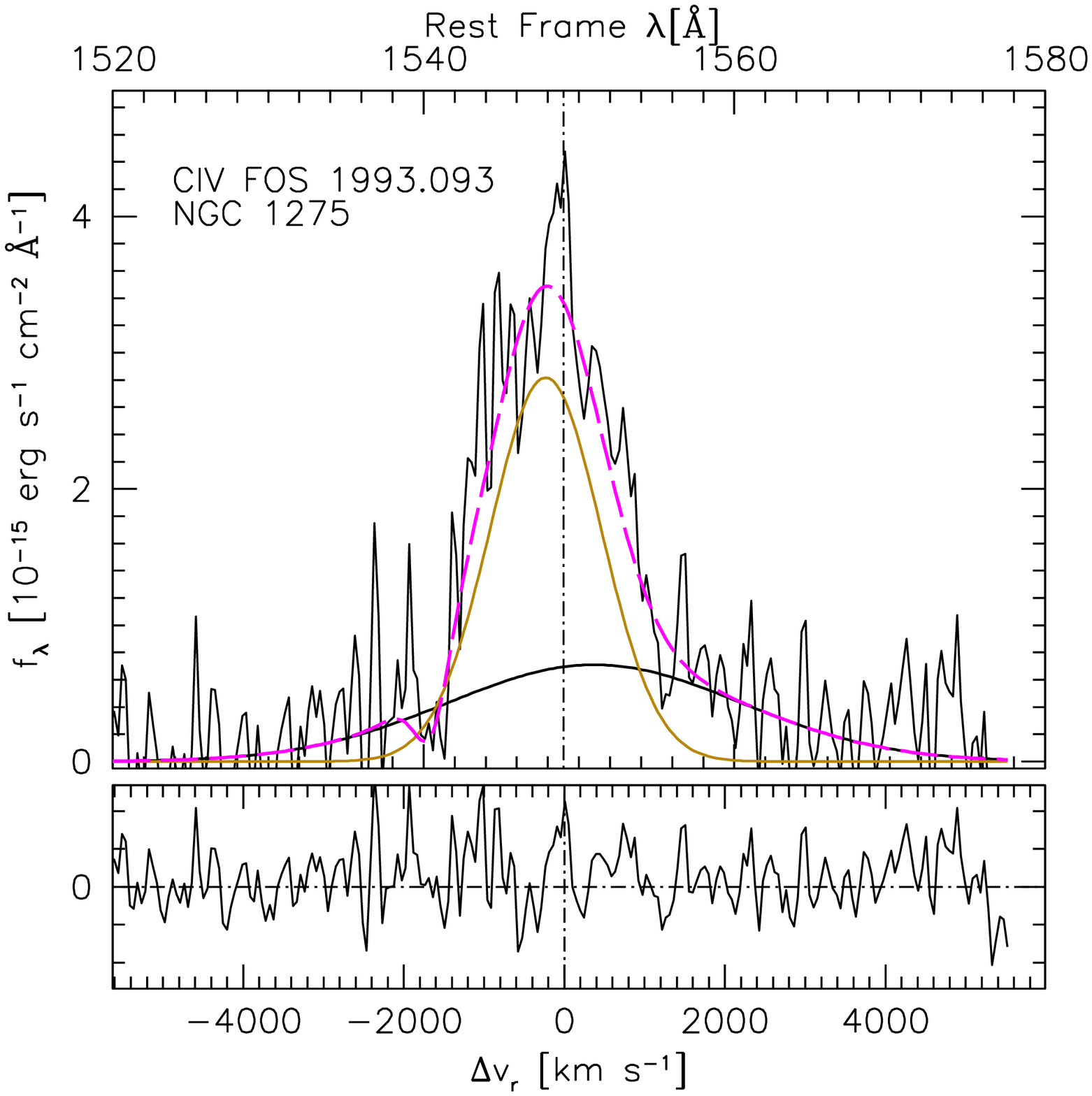}
\includegraphics[width=75 mm, angle= 0]{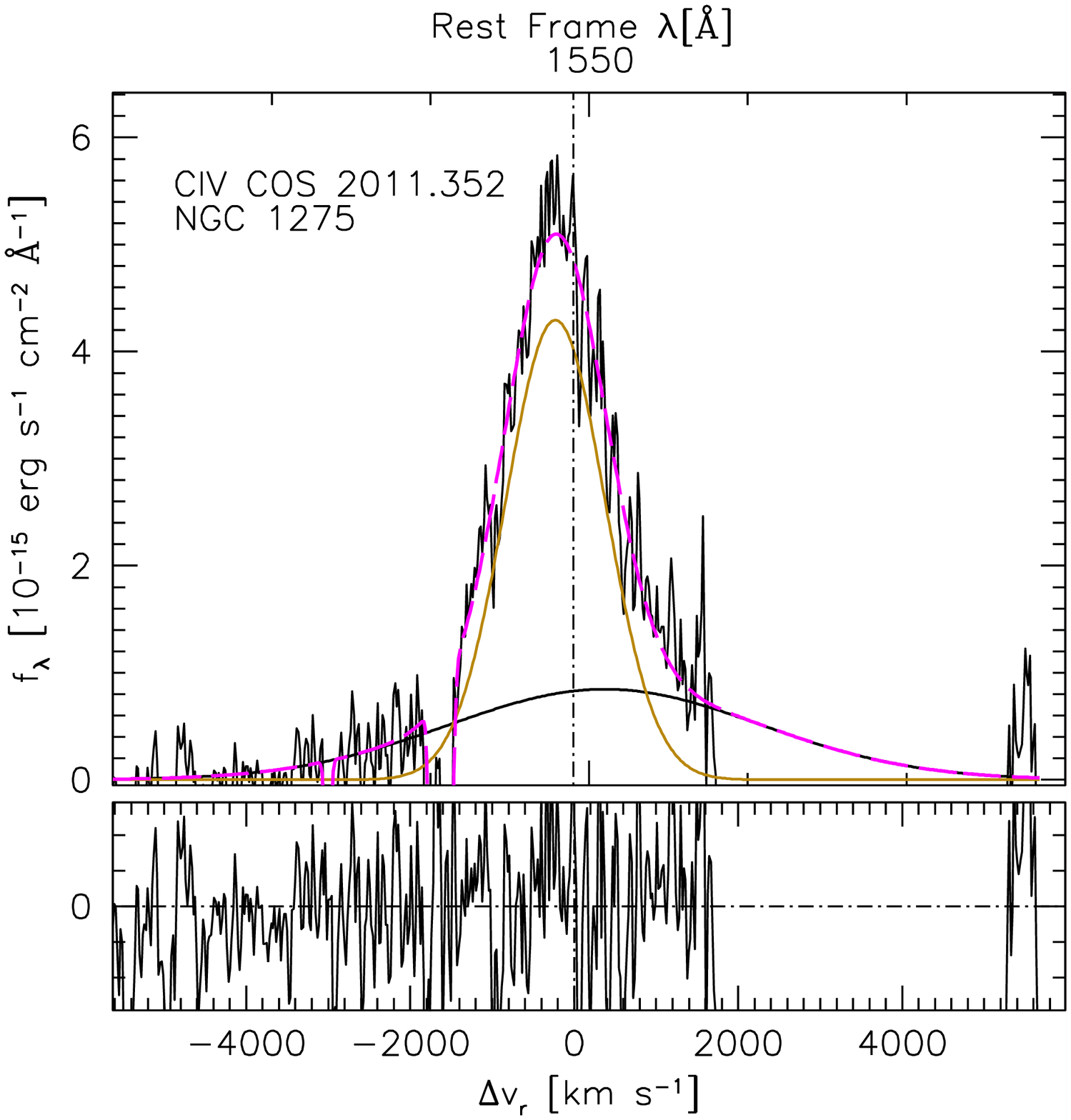}
\caption{The top frame illustrates that the widest plausible NL fit
to CIV in 2011.352 still results in a significant redwing residual.
The bottom frame fits the CIV profile with a NL+BEL in 1993.093
(left) and 2011.352 (right).}
\end{center}
\end{figure}
\section{The Nature of the Accretion Flow}The cause of the weak BELs and CIV in particular are a major
diagnostic of the central AGN. The weak BEL can arise from
\begin{enumerate}
\item a lack of BEL gas
\item intrinsic extinction
\item a weak ionizing continuum
\end{enumerate}
This section synthesizes broad band data and CLOUDY simulations in
order to discriminate between these three possibilities.
\subsection{The Ionizing Continuum}
In order to analyze these alternatives, we constrain the accretion
flow properties. This is difficult because the accretion flow is
under-luminous and is typically dominated in the optical band by the
high frequency synchrotron tail of the jet emission. The best chance
of finding the signature of the accretion flow is to look at high
frequency, where synchrotron cooling tends to lower the luminosity,
most importantly in a very low state. The 1993 HST FOS far UV
observation in Table 1 fits these criteria. The narrow slit is
advantageous for reducing background flux, but requires a careful
observation to avoid centering losses. There is an estimated
centering offset of 0.034" from the acquisition procedure
\citep{eva04}. Thus, we do not expect centering losses to affect the
absolute flux calibration. The spectral index of the continuum,
after Galactic de-reddening, $\alpha_{\lambda} \approx -1.5$
($F_{\lambda}\sim \lambda^{\alpha_{\lambda}}$), is typical of the
far UV continuum ($ 1100\AA < \lambda< 1600 \AA$) in low redshift,
low luminosity type 1 AGN based on HST spectra \citep{ste14}. The
spectral energy is $\lambda L_{\lambda}(1450 \AA) = 3.5 \times
10^{42}\rm{ergs/s}$. These numbers suggest a comparison to low
luminosity broad line Seyfert galaxies. Fortunately, there are low
luminosity Seyfert galaxies in the broad line reverberation sample.
The reverberation study of \citet{kas05} has a wealth of broadband
data (i.e., far UV, X-ray and H$\beta$ BEL luminosity). This sample
covers a large range of far UV luminosity that photo-ionizes the BEL
in the reverberation methodology. In order to ensure that we are
comparing to potentially similar objects, we choose a ``control"
subsample defined by an upper limit $\lambda L_{\lambda}(1450 \AA) <
10^{44}\rm{ergs/s}$, i.e. $\sim 30$ times stronger than 3C 84, in
order to study the nature of the ionizing continuum.
\par Figure 3 characterizes 3C 84 in the context of the
\citet{kas05} control sample. The left hand side plots the H$\beta$
BEL luminosity, $L(H\beta)$, versus the far UV luminosity, $\lambda
L_{\lambda}(1450 \AA)$. This is a good surrogate for the soft
ionizing continuum at $\lambda < 900\AA$. We added NGC 1275 to the
scatter plot by using the 1994 $L(H\beta)$ in Table 1, since it is
closest in time to the far UV observation. The left hand frame
indicates that $L(H\beta)/\lambda L_{\lambda}(1450 \AA)$ is typical
of what one expects for a low luminosity broad line Seyfert galaxy.
$L(H\beta)$ trends with $\lambda L_{\lambda}(1450 \AA)$ in Figure 3,
indicating that a weak photo-ionizing continuum is responsible for
the small $L(H\beta)$ for both the \citet{kas05} control sample and
NGC 1275. NGC 1275 is not an outlier in the $L(H\beta)$-$\lambda
L_{\lambda}(1450 \AA)$ scatter plane contrary to the expectation
that the weak BEL is a consequence of a paucity of BEL gas relative
to the control sample (condition 1, above).
\begin{figure}
\begin{center}
\includegraphics[width=75 mm, angle= 0]{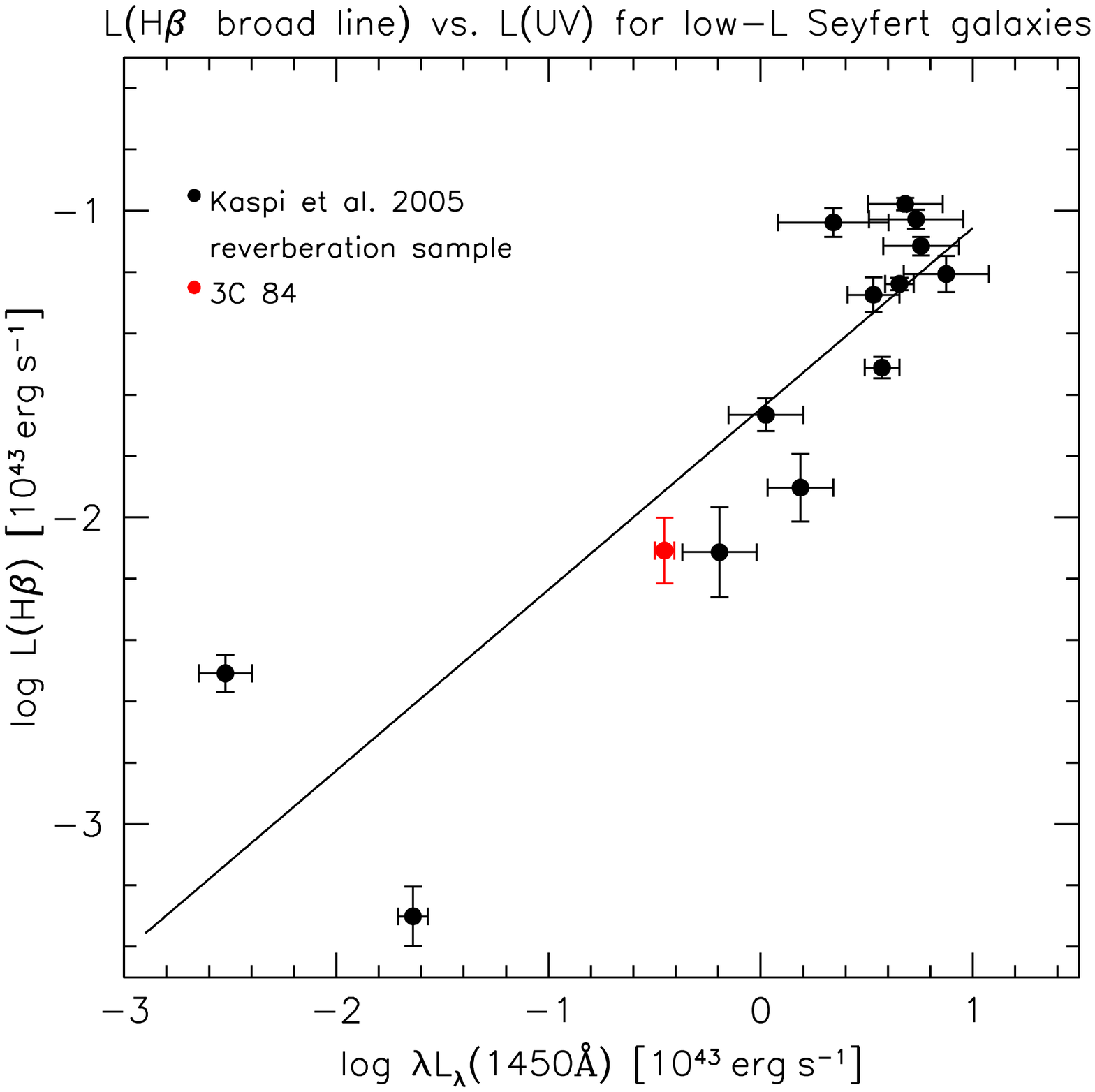}
\includegraphics[width=75 mm, angle= 0]{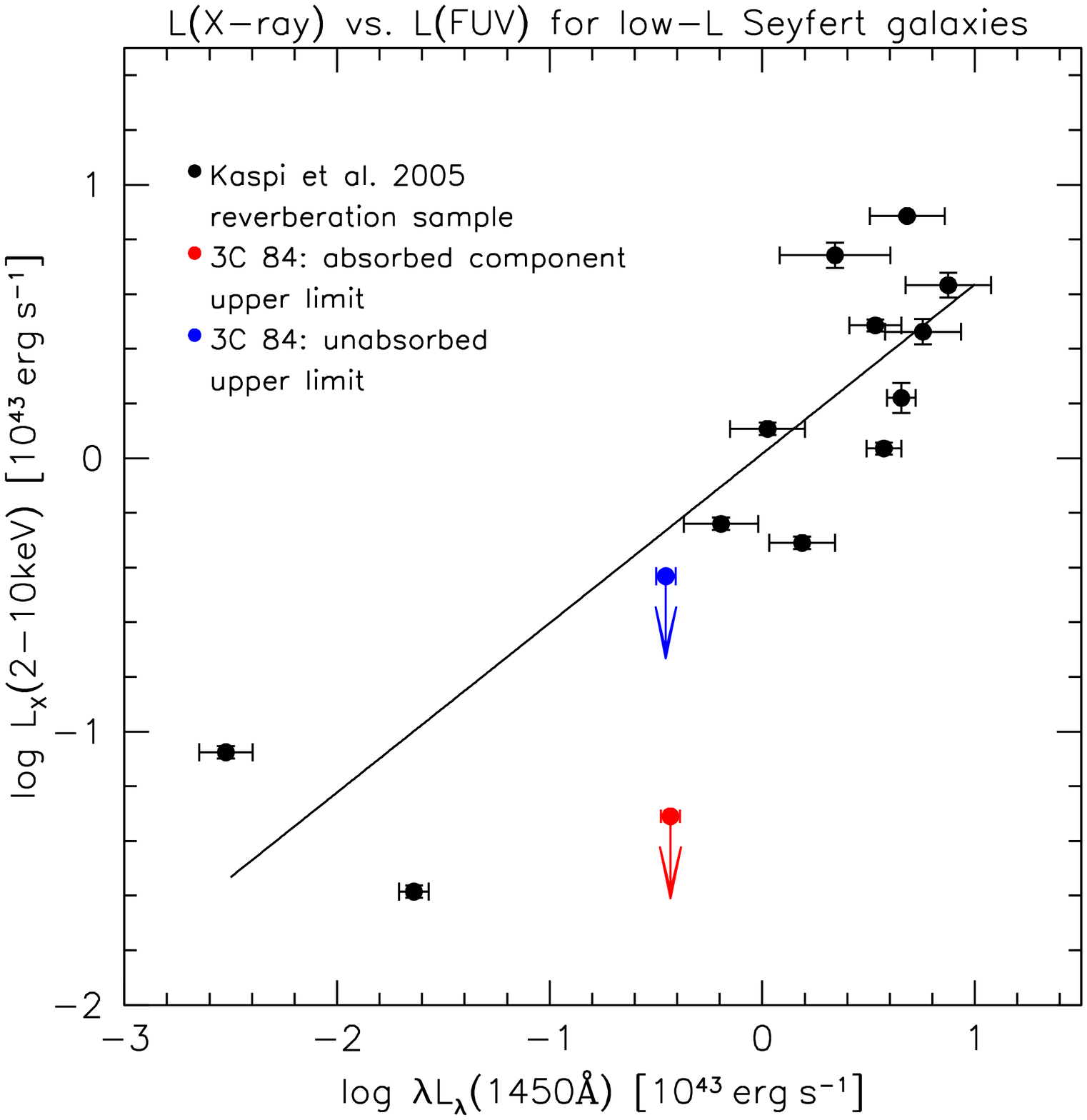}
\caption{Diagnostics of the photo-ionizing continuum. The left frame
shows that 3C 84 has a far UV-H$\beta$ ratio similar to that of weak
broad line Seyfert galaxies. The right frame indicates that the
ionizing continuum is softer than a typical broad line Seyfert
galaxy of similar far UV luminosity. The unabsorbed Chandra estimate
is a very loose upper bound as discussed in the text. The lines are
the best fit power laws to the \citet{kas05} data.}
\end{center}
\end{figure}
\par Next, we try to assess the hard ionizing continuum by looking at
the X-ray luminosity between 2 and 10 keV, $L_{2-10}$. The X-ray
luminosity of the accretion flow is not known, but there are some
useful upper limits. Chandra observations in 2002 are used to
estimate $L_{2-10}$ within $\sim$ 1.2" of the nucleus \citep{bal06}.
A short, off-axis, observation, $\sim$5 ks, that had a very low
count rate compared to other observations before 2006, 21 cts/s, was
utilized in order to minimize the effects of pile-up associated with
the nucleus. The spectrum is unabsorbed, and is likely primarily of
jet origin which consistently has a mm band luminosity $>10^{43}
\rm{ergs/s}$ and a gamma-ray luminosity $\sim 10^{44} \rm{ergs/s}$
\citep{abd10}. Thus, this is a very loose upper bound on $L_{2-10}$
from the accretion flow in the right hand frame of Figure 3. Another
method of estimating $L_{2-10}$ from accretion is to look for an
absorbed component of X-ray flux. This has been done based on XMM
data \citep{har09}. Their result is shown as an upper bound in the
right hand frame of Figure 3. This figure indicates that
$L_{2-10}/\lambda L_{\lambda}(1450 \AA)$ is much less than would be
expected for a low luminosity broad line Seyfert galaxy and contrary
to the basic prediction of an ADAF \citep{mah97}.
\par In summary, the two panels of Figure 3 and $\alpha_{\lambda} \approx -1.5$ in the far UV
seem to indicate a broad line Seyfert-like AGN with a typical soft
ionizing continuum and a weak hard ionizing continuum. Broad line
Seyfert-like AGN are believed to be a consequence of thermal
emission from optically thick, geometrically thin accretion
\citep{sun89}.
\begin{figure}
\begin{center}
\includegraphics[width=100 mm, angle= 0]{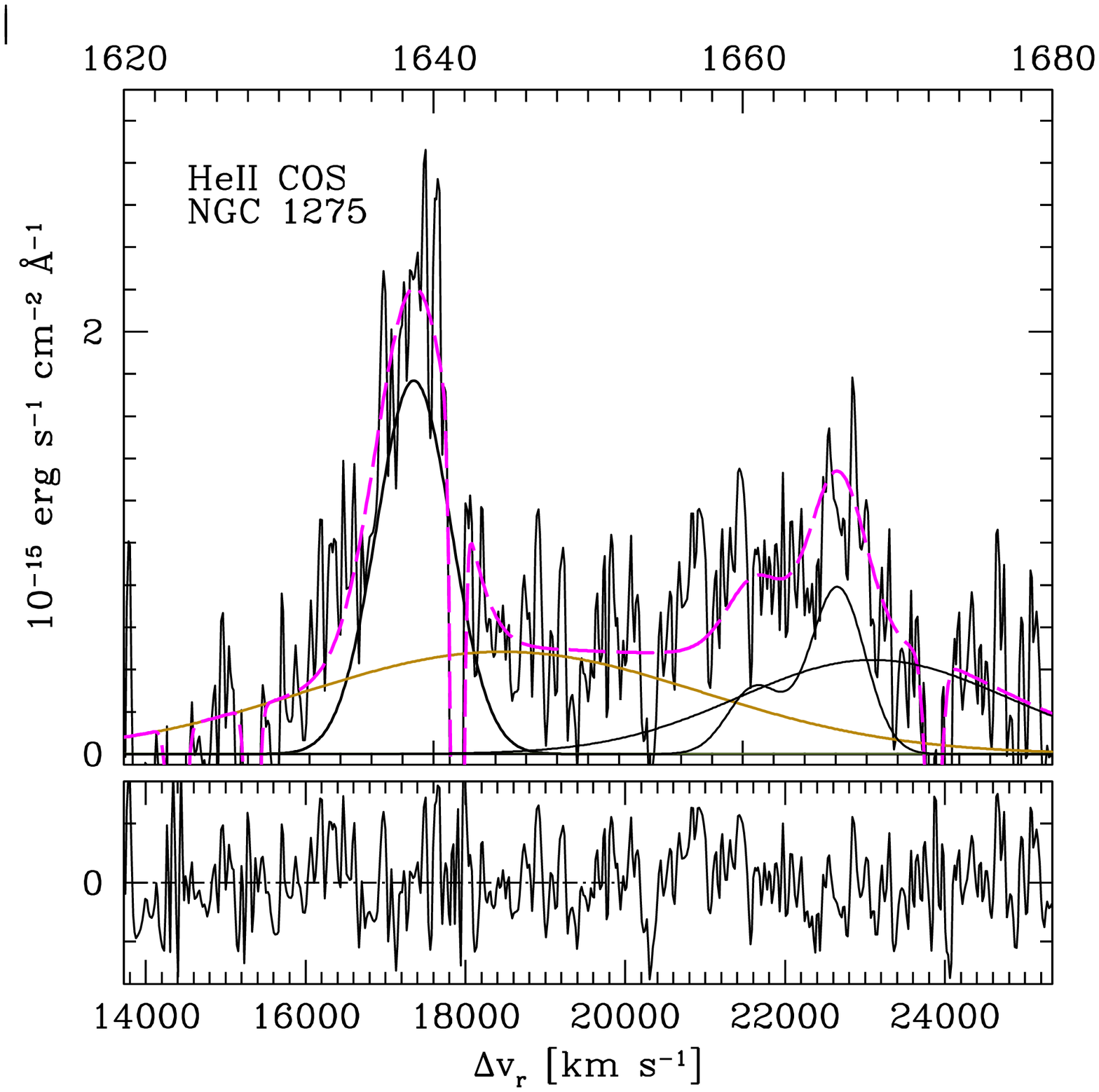}
\caption{Low ionization parameter simulations predict a strong
HeII$\lambda$1640 BEL relative to the nearby CIV BEL. The fit is
fairly complicated. There are narrow and broad (brown) HeII lines
and a broad AlII$\lambda$1671 line. In addition there are the two
NLs OIII]$\lambda$ 1666 and OIII]$\lambda$1663 that are fit in a
2.5:1 ratio. The estimate of the HeII BEL line strength is recorded
in Table 1.}
\end{center}
\end{figure}
\par Not only is the ionizing continuum weak based on
$R_{\rm{Edd}}$, but it is also soft. Both factors would conspire to
make the recombination rate fast relative to the photo-ionization
rate. Thus, less cooling occurs through high ionization state atomic
transitions. The weak ionizing continuum of 3C 84 is a viable
explanation of the low CIV/H$\beta$ BEL luminosity ratio. Even
though the BEL region is complicated by stratification and
anisotropy, simulations of individual clouds routinely show that
$L(CIV)$ decreases faster than $L(H\beta)$ as the ionizing continuum
becomes more dilute \citep{kor97}.
\subsection{CLOUDY Simulations} Table 1 provides some very strong
constraints of the BEL region. If we take into account, the
uncertainty in our estimates and some possible modest intrinsic
extinction that affects the UV more than the optical, we have the
following BEL constraints,
\begin{equation}
\frac{L(CIV)}{L(H\beta)} = 0.3-0.8\;, \,
\frac{L(P\alpha)}{L(H\beta)} =0.6 -1.1 \;, \,
\frac{L(FeII)}{L(H\beta)} < 0.3 \;, \, \frac{L(HeII\lambda
1640)}{L(CIV)} = 0.5 -1.0 \;.
\end{equation}
The details of the HeII fit to the 2003 COS data are shown in Figure
4. Note that the HeII/CIV ratio is unaffected by possible centering
issues that were noted in the footnote to Table 1. We can use these
ratios in order to explore the nature of the ionizing continuum and
the nature of the BEL gas.
\begin{figure}
\begin{center}
\includegraphics[width=100 mm, angle= 0]{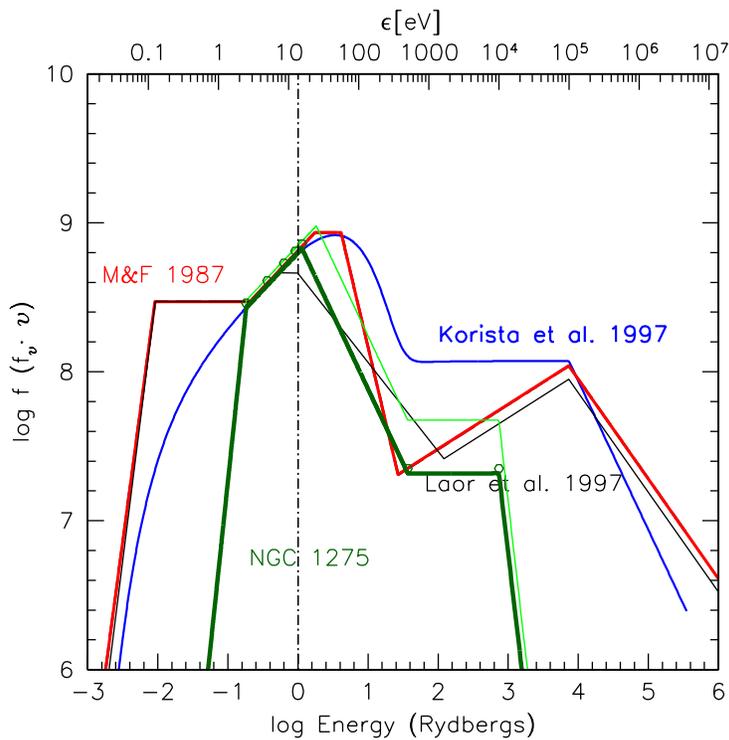}
\caption{Comparison of the ionizing continuum inferred for NGC 1275
from the last subsection compared to other popular ionizing continua
used in the literature as input spectra for CLOUDY simulations of
AGN BELs \citep{mat87,kor97,lao97}. The ionizing spectrum of NGC
1275 in green is softer.}
\end{center}
\end{figure}
\par We performed CLOUDY 13
single zone models (see Table 2) in order to achieve a qualitative
understanding of the constraints on the BEL and photo-ionizing
continuum that are consistent with the observations of NGC 1275
\citep{fer13}. Section 4.1 and Figure 3 describe our observed
constraints on the photo-ionizing continuum. Our estimated
photo-ionizing continuum has a spectral index $\alpha_{\nu} = 0.5$
(as observed in the far UV) from $5000 \AA$ to $800 \AA$. The
spectral break point at $800 \AA$ is fairly typical of other low
luminosity Seyfert galaxies \citep{ste14}. Beyond the spectral
break, we choose $\alpha_{\nu} = 2$ down to the soft X-ray band at
0.5 keV. The spectral index is chosen to be $\alpha_{\nu} = 1.0$ in
the X-ray band (see Figure 5).
\begin{table}
\caption{CLOUDY Simulations: NGC 1275}
{\tiny\begin{tabular}{ccccccccccc} \tableline \rule{0mm}{3mm}
1 &  2  &  3 &  4  & 5  & 6 & 7 & 8 & 9 & 10 & 11 \\
\tableline\rule{0mm}{3mm}
 Z & $\log{n}$ &  $\log{N_{H}}$ & $\log{r}$   & $\log{U}$ & $\log{L(H\beta)}$  & CIV/$H\beta$ &  P$\alpha$/$H\beta$ & HeII/CIV & FeII/$H\beta$ & Conformance \\
\tableline \rule{0mm}{3mm}
Metalicity & Number  &  Column & Radius & Ionization & Luminosity &  & &  &  &  \\
 &  Density &  Density &    & Parameter  &  & Target & Target & Target & Target   &   \\
  &  $\rm{cm}{-3}$ & $\rm{cm}{-2}$ & cm &  & erg/sec & 0.3 - 0.8  & 0.6 - 1.1  & 0.5 - 1.0 & $<0.3$&   \\
\tableline \rule{0mm}{3mm}
0.1 & 9.75  & 22.7 & 16.5 & -1.24  & 40.76 & 1.63 & 0.99 & 0.44 & 0.03 & No\\
0.1 & 9.75  & 23.3 & 16.5 & -1.24  & 40.80 & 1.47 & 1.09 & 0.43 & 0.06 & No\\
0.1 & 9.75  & 22.7 & 16.75& -1.74  & 40.76 & 1.63 & 0.99 & 0.44 & 0.03 & No\\
0.1 & 9.75  & 23.3 & 16.75& -1.74  & 40.80 & 1.47 & 1.09 & 0.43 & 0.06 & No\\
0.1 & 9.75  & 22.7 & 17.0 & -2.24  & 40.76 & 1.63 & 0.99 & 0.44 & 0.03 & No\\
0.1 & 9.75  & 23.3 & 17.0 & -2.24  & 40.80 & 1.47 & 1.09 & 0.43 & 0.06 & No\\
0.1 & 10.0  & 22.7 & 16.5 & -1.49  & 40.81 & 0.63 & 0.92 & 0.99 & 0.02 & Yes\\
0.1 & 10.0  & 23.3 & 16.5 & -1.49  & 40.84 & 0.59 & 1.01 & 0.97 & 0.05 & Yes\\
0.1 & 10.0  & 22.7 & 16.75& -1.99  & 40.81 & 0.63 & 0.91 & 0.99 & 0.02 & Yes\\
0.1 & 10.0  & 23.3 & 16.75& -1.99  & 40.84 & 0.59 & 1.01 & 0.97 & 0.05 & Yes\\
0.1 & 10.0  & 22.7 & 17.0 & -2.49  & 40.81 & 0.63 & 0.91 & 0.99 & 0.02 & Yes\\
0.1 & 10.0  & 23.3 & 17.0 & -2.49  & 40.84 & 0.59 & 1.01 & 0.97 & 0.05 & Yes\\
0.1 & 10.25 & 22.7 & 16.5 & -1.74  & 40.87 & 0.20 & 0.84 & 2.64 & 0.02 & No\\
0.1 & 10.25 & 23.3 & 16.5 & -1.74  & 40.89 & 0.19 & 0.93 & 2.59 & 0.04 & No\\
0.1 & 10.25 & 22.7 & 16.75& -2.24  & 40.87 & 0.20 & 0.84 & 2.64 & 0.02 & No\\
0.1 & 10.25 & 23.3 & 16.75& -2.24  & 40.89 & 0.19 & 0.93 & 2.59 & 0.04 & No\\
0.1 & 10.25 & 22.7 & 17.0 & -2.74  & 40.87 & 0.20 & 0.84 & 2.64 & 0.02 & No\\
0.1 & 10.25 & 23.3 & 17.0 & -2.74  & 40.89 & 0.19 & 0.93 & 2.59 & 0.04 & No\\
0.1 & 10.5  & 22.7 & 16.5 & -1.99  & 40.93 & 0.06 & 0.77 & 8.22 & 0.02 & No\\
0.1 & 10.5  & 23.3 & 16.5 & -1.99  & 40.94 & 0.05 & 0.87 & 8.08 & 0.03 & No\\
0.1 & 10.5  & 22.7 & 16.75& -2.49  & 40.93 & 0.06 & 0.77 & 8.22 & 0.02 & No\\
0.1 & 10.5  & 23.3 & 16.75& -2.49  & 40.94 & 0.05 & 0.87 & 8.08 & 0.03 & No\\
0.1 & 10.5  & 22.7 & 17.0 & -2.99  & 40.93 & 0.06 & 0.77 & 8.22 & 0.02 & No\\
0.1 & 10.5  & 23.3 & 17.0 & -2.99  & 40.94 & 0.05 & 0.87 & 8.08 & 0.03 & No\\
\tableline \rule{0mm}{3mm}
1.0 & 9.75  & 22.7 & 16.5 & -1.24  & 40.75 & 11.8 & 0.33 & 0.07 & 0.04 & No\\
1.0 & 9.75  & 23.3 & 16.5 & -1.24  & 40.83 & 9.77 & 0.35 & 0.07 & 0.07 & No\\
1.0 & 9.75  & 22.7 & 16.75& -1.74  & 40.82 & 7.67 & 0.60 & 0.08 & 0.08 & No\\
1.0 & 9.75  & 23.3 & 16.75& -1.74  & 40.89 & 6.49 & 0.54 & 0.08 & 0.11 & No\\
1.0 & 9.75  & 22.7 & 17.0 & -2.24  & 40.76 & 3.85 & 1.02 & 0.18 & 0.19 & No\\
1.0 & 9.75  & 23.3 & 17.0 & -2.24  & 40.78 & 3.65 & 1.05 & 0.18 & 0.27 & No\\
1.0 & 10.0  & 22.7 & 16.5 & -1.49  & 40.71 & 11.2 & 0.32 & 0.07 & 0.05 & No\\
1.0 & 10.0  & 23.3 & 16.5 & -1.49  & 40.77 & 9.72 & 0.36 & 0.07 & 0.08 & No\\
1.0 & 10.0  & 22.7 & 16.75& -1.99  & 40.85 & 5.20 & 0.52 & 0.11 & 0.07 & No\\
1.0 & 10.0  & 23.3 & 16.75& -1.99  & 40.88 & 4.84 & 0.56 & 0.11 & 0.11 & No\\
1.0 & 10.0  & 22.7 & 17.0 & -2.49  & 40.79 & 1.91 & 0.97 & 0.32 & 0.22 & No\\
1.0 & 10.0  & 23.3 & 17.0 & -2.49  & 40.81 & 1.84 & 1.00 & 0.32 & 0.22 & No\\
1.0 & 10.25 & 22.7 & 16.5 & -1.74  & 40.69 & 9.33 & 0.35 & 0.09 & 0.06 & No\\
1.0 & 10.25 & 23.3 & 16.5 & -1.74  & 40.73 & 8.40 & 0.34 & 0.09 & 0.09 & No\\
1.0 & 10.25 & 22.7 & 16.75& -2.24  & 40.81 & 3.55 & 0.55 & 0.17 & 0.08 & No\\
1.0 & 10.25 & 23.3 & 16.75& -2.24  & 40.86 & 3.15 & 0.54 & 0.17 & 0.11 & No\\
1.0 & 10.25 & 22.7 & 17.0 & -2.74  & 40.83 & 0.76 & 0.91 & 0.73 & 0.12 & Yes\\
1.0 & 10.25 & 23.3 & 17.0 & -2.74  & 40.85 & 0.73 & 0.94 & 0.73 & 0.18 & Yes\\
1.0 & 10.5  & 22.7 & 16.5 & -1.99  & 40.66 & 7.09 & 0.34 & 0.13 & 0.07 & No\\
1.0 & 10.5  & 23.3 & 16.5 & -1.99  & 40.68 & 6.86 & 0.41 & 0.13 & 0.11 & No\\
1.0 & 10.5  & 22.7 & 16.75& -2.49  & 40.82 & 1.85 & 0.54 & 0.32 & 0.08 & No\\
1.0 & 10.5  & 23.3 & 16.75& -2.49  & 40.83 & 1.81 & 0.54 & 0.32 & 0.12 & No\\
1.0 & 10.5  & 22.7 & 17.0 & -2.99  & 40.88 & 0.23 & 0.84 & 2.09 & 0.10 & No\\
1.0 & 10.5  & 23.3 & 17.0 & -2.99  & 40.89 & 0.22 & 0.84 & 2.09 & 0.15 & No\\
\end{tabular}}
\end{table}
The results of our CLOUDY simulations are presented in Table 2. All
the simulations have a similar L(H$\beta$) and a covering fraction
of 1 in an open geometry in order to explore trends that are
conducive to reproducing the observed line ratios. The ionization
parameter, $U$, is the number density of ionizing photons for
hydrogen (number density of photons with $\lambda < 912 \AA$)
divided by the hydrogen number density, $n$ in column (2), in the
single zone model. We identify two regions of parameter space that
conform to the constraints imposed by observation. The first
possibility is a very low ionization parameter $\log{U} \approx
-2.5$ with solar metallicity ($Z=1$) at large distances from the
ionizing source, $\lesssim 10^{17} \rm{cm}$. The other plausible
region of parameter space has a slightly low ionization parameter
$\log{U} \approx -1.5$ and a loosely constrained distance from the
source, $3 \times 10^{16} \rm{cm} - 10^{17} \rm{cm}$, but requires a
very low metallicity. The first alternative is what was expected
based on the discussion at the end of Section 4.1.
\par These simulations show that the ionizing continuum of NGC 1275
is consistent with the unusual BEL line ratios and the low BEL
luminosity. In particular, it explains the weak CIV BEL luminosity
relative to the H$\beta$ luminosity as well as the inordinately
strong HeII$\lambda$1640 BEL line strength relative to the CIV line
strength.

In order to distinguish between the $Z=0.1$ and $Z=1.0$ solutions,
we explore the Eddington ratio for more constraints on the system.

\subsection{Eddington Ratio}
The possibility of an optically thick thermal accretion flow is
interesting considering the low luminosity. We assume the far UV
power law extends from $10000 \AA$ to as far down $500 - 800 \AA$,
at most, before breaking sharply down to the X-ray continuum
\citep{ste14}. The bolometric luminosity of the accretion flow would
only be $L_{\rm{bol}} =1-2 \times 10^{43} \rm{ergs/s}$. This
liberally assumes a range of X-ray luminosity from negligible to
roughly equal to the extrapolated UV continuum. In order to quantify
this we note that the central black hole mass is estimated as
$M_{bh}\approx 8 \times 10^{8} M_{\odot}$ based on the NICMOS bulge
luminosity estimators \citep{don07,mar03}. Similarly, gas
kinematical estimates find $ 4 \times 10^{8}<M_{bh}/M_{\odot} <2
\times 10^{9} $ \citep{sch13}. With this mass estimate, the
Eddington rate, $0.00004< R_{\rm{Edd}} < 0.0006$, is low compared to
low redshift Seyfert 1 galaxies that typically have Eddington rates
on the order of a few percent \citep{sun89}. We want to explore this
circumstance in terms of the CLOUDY models, but we need more
constraints.
\subsection{Redshift of the BEL Region} Table 2 provides a range of
BEL distances from the ionization source that are consistent with
the weak, soft ionizing continuum and the BEL line ratios and line
strengths. We have an extra constraint that arises based on the
estimates of $M_{bh}$. We note that in our fits, the BELs are
redshifted relative to the systemic velocity of NGC 1275. However,
this redshift is modest, $\sim 500 \rm{km/sec}$, compared to other
high (central black hole) mass broad line objects \citep{sul07}.
Knowing this value to more than a 50\% uncertainty is beyond the
ability of our methods applied to these weak BELs. The gravitational
redshift in terms of velocity and geometrized black hole mass, $M$,
is \citep{cor98}
\begin{equation}
v_{\rm{redshift}}/c \approx M/r\;,\quad M \approx
\frac{M_{bh}}{10^{9}M_{\odot}}1.4\times 10^{14} \rm{cm}.
\end{equation}
Motional Doppler shifts can be of comparable magnitude, but can only
be estimated based on knowing the velocity distribution of the BEL
gas and line of sight. These are unknown and their inclusion would
be speculative. We note that a pure transverse Doppler shift from
transverse Keplerian motion adds a factor of 1.5 to Equation (2)
\citep{cor98}. The gravitational redshift can also be partially
canceled by an out-flowing BEL gas and there is a strong jet in 3C
84 that can conceivably induce an outflow in the BEL region. All we
know is that the gravitational redshift is much lower than the 1000
km/sec - 2000 km/sec redshifts typical of sources with large central
black holes and small Eddington ratios \citep{sul07}. The modest
observed BEL redshifts and Equation (2) crudely limits the range of
allowable distances from the photo-ionization source to $> \sim 400-
500 M$. We can relate this constraint to the simulations in Table 2
\begin{equation}
\log{r} > 16.75\;, \; \frac{M_{bh}}{10^{9}M_{\odot}} =1\;, \quad
\log{r}> 16.5\;, \quad \frac{M_{bh}}{10^{9}M_{\odot}} =0.5
\end{equation}

\subsection{Virial Mass Estimates} A related issue are the single epoch virial estimates of
$M_{bh}$. We are proposing that NGC 1275 is similar to the low
luminosity sample of \citep{kas05}, but $M_{bh}$ is clearly much
larger than these reverberation sources. So, we explore the virial
mass estimators in order to get more restrictions on the location of
the BEL region. We cannot use the optical continuum luminosity based
estimators since the optical flux is dominated by the synchrotron
jet and not the accretion flow emission. Thus, we need to use
L(H$\beta$) as a surrogate for the ionizing continuum. The formula
of \citet{gre05,wuu04} gives us the type of estimate that we can
utilize,
\begin{equation}
\frac{M_{bh}}{M_{\odot}} = 3.6 \times 10^{6}
\left[\frac{L(H\beta)}{10^{42} \,\rm{erg/s}}\right]^{0.56} \left[
\frac{\rm{FWHM}}{1000 \rm{km/s}}\right]^{2}\;.
\end{equation}
It is very important to note that the reverberation sources at
$L(H\beta)< 5 \times 10^{41}\rm{ergs/sec}$ do not correlate strongly
with the reverberation radius \citep{wuu04,kas05}. Thus, Equation
(4) has much larger uncertainty at the low end of its range than at
its high end. Using the results of Table 1 we estimate
$\frac{M_{bh}}{M_{\odot}} \approx 1.5-3.5 \times 10^{7}$. This is a
factor or 10 - 100 less than the estimates in Section 4.3.

\par In order to understand this discrepancy, we consider the BL-Lac
nature of NGC 1275 \citep{ver78}. The BL-Lac properties of 3C 84
have been noted in connection with the synchrotron optical emission
that hides the accretion generated continuum. The BL-Lac aspect can
be very pronounced with optical polarization that changes
dramatically in amplitude and position angle, with the largest
optical polarization reaching 6\% \citep{ang80}. Such extreme
blazar-like properties suggest a small line of sight to the jet axis
\citep{lin85}. We qualify this statement by noting that the evidence
does not support extreme BL-Lac behavior. The range of polarization
is 1-6 \% in \citet{ang80} and more extensive polarization data
\footnote{from http://www.bu.edu/blazars/VLBAproject.html} covering
the time period 2011-2018 indicates an optical polarization that is
usually below 2\% and extremely rare (2) instances of polarization
$>3\%$ were reported. This suggests a slightly off angle BL-Lac. The
H$\beta$ BEL is considered to be rotating gas in a flat
``pancake-like" region in which the normal to the BEL disk is
parallel to the jet axis \citep{bro86}. Thus, the FWHM that appears
in Equation (4) depends on the line of sight. For polar lines of
sight, Equation (4) will under estimate $M_{bh}$. In the virial mass
estimation,
\begin{equation}
M_{bh} = \frac{R_{\rm{BLR}}v_{\rm{BLR}}^{2}}{\rm{G}}\;,
\end{equation}
where G is the gravitational constant, $R_{\rm{BLR}}$ is the orbital
radius of the BLR (broad line region) and $v_{\rm{BLR}}$ is the
velocity of the BEL gas. In order to relate $v_{\rm{BLR}}$ to an
observed quantity, one defines the de-projection factor, $f$,
\begin{equation}
v_{\rm{BLR}} = f \rm{FWHM}\;.
\end{equation}
Equations (5) and (6) indicate that $M_{bh} \propto f^{2}$. The
de-projection factor has been estimated for various classes of
objects that are believed to be differentiated by a line of sight
(LOS) \citep{ant93,dec11}. The method of \citet{dec11} was to
estimate $M_{bh}$ from the bulge luminosity of the host galaxy.
Using this estimate to set the value of $M_{bh}$ in the virial
formula they were able to estimate $f$ for various classes of
objects,
\begin{equation}
f_{\rm{isotropic}}= \frac{\sqrt{3}}{2}\;, \quad
f_{\rm{quasars}}=2.0\pm0.3\;,\quad
f_{\rm{blazars}}=5.6\pm1.3\;,\quad f_{\rm{BL-Lacs}}=6.9\pm 2.3 \;.
\end{equation}
Equation (4) is derived based on assuming an isotropic distribution
of BEL gas velocity. Thus, we adopt a correction factor for the
estimate in Equation (4) of
$(f_{\rm{BL-Lacs}}/f_{\rm{isotropic}})^{2}$ for BL-Lac orientations
such as the one that exists in 3C 84. Taking the nominal value of
\citet{dec11} in Equation (7), we expect a correction factor of 64.
The orientation corrected central black hole mass estimate based on
the data in Table 1 yields $\frac{M_{bh}}{M_{\odot}} \approx 1.0-
3.2\times 10^{9}$. Alternatively, if we use the blazar correction of
42 associated with the nominal value of $f_{\rm{blazars}}$ in
Equation (7) instead, we get $\frac{M_{bh}}{M_{\odot}} \approx 0.6-
1.5\times 10^{9}$.
\par These corrections seem very extreme and we wish to see how they
relate to our CLOUDY fits and the redshift constraints on the
distance to the BEL region in Equation (3). Assuming a FWHM = 5000
km/s for H$\beta$ from Table 1 and using Equation (5)
\begin{eqnarray}
&&\log{r} = 16.75\;,\;\frac{M_{bh}}{10^{9}M_{\odot}}=1\;,\quad
f=3.1\;,\quad \left[\frac{f}{f_{\rm{isotropic}}}\right]^{2} = 12.8\\
 &&\log{r}= 16.5\;,\; \frac{M_{bh}}{10^{9}M_{\odot}} =0.5\;,
\quad f = 2.9 \;,\quad \left[\frac{f}{f_{\rm{isotropic}}}\right]^{2}
= 11.2\;.
\end{eqnarray}
Using the line of sight correction factors from Equations (8) and
(9) in the virial mass estimator Equation (4) one gets
\begin{eqnarray}
&&\log{r} = 16.75\;,\;\frac{M_{bh}}{10^{9}M_{\odot}}=1\;,\quad
f=3.1\;,\quad \frac{M_{\rm{virial}}}{M_{bh}} \approx 0.19-0.44\\
 &&\log{r} = 16.5\;,\; \frac{M_{bh}}{10^{9}M_{\odot}} =0.5\;,
\quad f = 2.9 \;,\quad \frac{M_{\rm{virial}}}{M_{bh}} \approx
0.33-0.78\;.
\end{eqnarray}
Considering the uncertainty in the virial estimate, especially at
the low luminosity end, Equations (10) and (11) are within
acceptable agreement once the LOS corrections were
applied.
\subsection{Mitigating the Tension Between Constraints}
Equations (1) - (4) combined with Table 2 provide many constraints
on the photo-ionization BEL system.
\begin{enumerate}
\item There are the four line ratios in Equation (1).
\item There are the compatible distances from the source in Table 2
that are associated with these ratios in a simple single zone model.
\item There are the constraints on distances from the source based on
the line redshifts in Equations (2) and (3).
\item There is the 1-2 order of magnitude under estimate of the
central black hole mass from the virial formula in Equation (4).
\item There is the BL-Lac nature of the source, perhaps slightly off
angle
\end{enumerate}
The strategy in a complicated system of constraints that are only
estimates is to find solutions that fully or partially mitigate the
tension between them. We find that this tension is reduced by a
slightly off angle blazar correction applied to the virial mass
estimate near the minimal allowed distances in the CLOUDY models.
The tension can be relieved further with small amounts of intrinsic
extinction and possibly low metallicity. We note that if the
distance to the BEL is $\log{r} < 16.75$ low metallicity is required
to satisfy the line ratio constraints in Equation (1) based on the
CLOUDY simulations. We consider $\log{r} = 16.75$ to be possibly
most viable. It conforms with $Z=0.1$ and is close to being
conforming with $Z=1.0$ in Table 2. It might be possible to fine
tune a solution with $Z=1.0$ in the over simplified model
(especially with some intrinsic extinction). We consider this the
best choice for mitigating the tension between constraints based on
the following:
\begin{enumerate}
\item It is close to reconciling the four line ratios in Equation (1) and the simple
single zone CLOUDY model with $Z=1.0$ within the coarse resolution
of Table 2. It does conform with $Z=0.1$.
\item For $M_{bh}=10^{9} M_{\odot}$, the gravitational redshift is only 750 km/sec.
\item The f value of 3.1 in equation (8) is intermediate between a quasar and a blazar
based on equation (7). Consistent with the modest BL-Lac behavior
\item The line of sight correction reconciles the viral black hole
estimate with $ M_{bh}\approx 10^{9} M_{\odot}$ within the viral
estimators uncertainty.
\end{enumerate}

In summary, the CLOUDY simulations in Table 2 indicate that a BEL
region with a number density of $n \sim 1-3 \times 10^{10}\,
\rm{cm}^{-3}$ located $\sim 5-10 \times 10^{16}$ cm that is
photo-ionized by the thermal emission from an optically thick
accretion flow is consistent with the BEL line ratios, the line
strengths and the velocity shifts relative to the systemic redshift
of 3C84. Furthermore, if the LOS is offset slightly from the normal
to planar rotation of the low ionizations lines then the virial mass
estimate of $M_{bh}$ conforms with bulge luminosity and kinematic
estimates. Integrating this information with the CLOUDY results, we
concluded that $r \approx 5\times 10^{16} \rm{cm}$ is most viable.
The mitigation of tension amongst constraints described in this
section reinforces the finding that 3C 84 has a BEL region that is
photo-ionized by the emission from an optically thick accretion flow
similar to the low luminosity Seyfert galaxies in the reverberation
sample \citep{kas05}.
\par The identification of a BL-Lac type jet
within the core of 3C 84 is not a trivial interpretation. The
interior jet geometry is far more complex than simple jet models
predict \citep{gio18}. The pc scale structure in terms of jet speed,
counter jet to jet flux ratios and free-free absorption strongly
indicate a jet orientation much closer to the sky plane than a polar
BL-Lac orientation \citep{wal00,fuj17}. Thus, the interior jet must
change orientation, drastically, from a  BL-Lac line of sight to an
oblique line of sight on very small scales. Possible evidence of
this is found in Figure 2 of \citet{gio18}, the inner jet
drastically changes direction and morphology $\sim 0.1$ pc from the
core.

\subsection{Long Term Time Evolution of the Accretion Flow}
We note that the current $L_{\rm{bol}}$ estimated in Section 4.3 is
much lower than would be inferred from [OIII], [OII], and the Mid-IR
(in order of increasing distance from the photo-ionizing source).
Based on the relative strength of the high ionization lines, the NL
region seems to be photo-ionized. From Table 1 we find
$L([OIII]\lambda 5007)$/L(H$\beta$(NL))$\sim 5-6$,
L([OIII])/L([OII])$\approx 2$. $L_{\rm{bol}}$ can be estimated from
$L([OIII])$, $L([OII])$, $\lambda L_{\lambda}(\lambda= 12\mu)$ and
$\lambda L_{\lambda}(\lambda= 25\mu)$ \citep{ste12,wil99,spi95},
\begin{eqnarray}
&& L_{\rm{bol}} = 4000 \left[\frac{L([OIII])}{10^{43}
\rm{ergs/s}}\right]^{1.39}\left[10^{43}\right] \rm{ergs/s} \;,\\
&& L_{\rm{bol}} \approx 5000 \, L([OII]) \;,\\
&& \log[L_{\rm{bol}}] = 0.942\log[\lambda L_{\lambda}(\lambda=
12\mu)] + 3.642 \;,\\
&& \log[L_{\rm{bol}}] = 0.837\log[\lambda L_{\lambda}(\lambda=
25\mu)] + 8.263 \;.
\end{eqnarray}
The 12$\mu$ (30$\mu$) excess over a synchrotron power law has a
luminosity of $1.58 \times 10^{44}$ erg/s ($2.44 \times 10^{44}$
erg/s) which has been ascribed to dust heated by the AGN
\citep{lei09}. This corresponds to $L_{\rm{bol}} \approx 2-2\times
10^{45} \rm{ergs/s}$ \citep{spi95}. For the values of [OIII] and
[OII] in Table 1, $L_{\rm{bol}}\approx 7.1\times 10^{44}
\rm{ergs/s}$ and $L_{\rm{bol}}\approx1.3\times 10^{45} \rm{ergs/s}$,
with the bolometric corrections in \citet{ste12,wil99},
respectively. Apparently, the accretion flow has decreased in
luminosity relative to its value many decades in the past.
Presently, $L_{\rm{bol}}$ is $\sim 0.01$ of the value imprinted in
the surrounding molecular gas. The size of the molecular dust based
on subarcsecond Mid-IR imaging is restricted to $<129$ pc
\citep{asm14}. Thus, the fading of the AGN continuum has likely been
occurring on a time frame of many decades to 300 years.

\section{The Jet-Accretion Flow Connection}
In order to establish a connection between the accretion flow state
and the jet power, $Q(t)$, requires the identification of observable
quantities that represent $L_{\rm{bol}}$ and $Q(t)$. In the last
section, it was argued that the H$\beta$ BEL luminosity is related
to the strength of the ionizing continuum as in other broad line
Seyfert galaxies \citep{gil03}. Thus, one of the main goals of this
section is to find a surrogate for $Q(t)$,
\subsection{A Surrogate for Jet Power} It is far from trivial to
find a surrogate for the jet power of a blazar-like jet. Although
numerous modeling schemes exist to accomplish this (eg.
\citet{ghi10}), the process is highly dependent on the degree of
Doppler enhancement of the observed flux density. The Doppler
factor, $\delta$, is given in terms of $\Gamma$, the Lorentz factor
of the outflow; $\beta$, the three velocity of the outflow and the
angle of propagation to the line of sight, $\theta$;
$\delta=1/[\Gamma(1-\beta\cos{\theta})]$ \citep{lin85}. The Doppler
enhancement of the luminosity of an unresolved source is
$\delta^{4}$ and can range from a factor of a ten to a factor of few
tens of thousands \citep{lin85,tin05}. The unresolved source Doppler
enhancement is particularly relevant to attempts to model the
broadband blazar jet emission. Due to the level of complexity these
models are generally reduced to single zone spherical models, the
mathematical analog of an unresolved source \citep{ghi10,boe13}. One
can try to constrain this Doppler factor by making assumptions on
the line of sight and the bulk flow of the jet. However, the errors
of such methods are also exaggerated by Doppler enhancement, so the
method can yield considerable scatter as a result of uncertainty and
random fluctuations in the line of sight and Lorentz factor
\citep{tin05,wil99}. In summary, the flux density is not a reliable
method of tracking the intrinsic change in jet power of a blazar
jet.
\par Fortunately, 3C 84 is not typical in this regard. The jet shows
order unity changes in mm flux on time scales of weeks as expected
for a blazar. This flickering is superimposed on a larger amplitude
long term variability time scale of many years to decades (as we
show in the data presented in this section). We postulate that the
long term background flux variation of the core can be used as a
surrogate for $Q(t)$ after the flickering is averaged out. Yet, we
do not believe that it can be used to determine $Q(t)$
quantitatively in a reliable manner. We ultimately substantiate this
postulate in Section 5.2 (with certain caveats), by revealing
evidence that the long term mm flux varies in consort with the
ionizing continuum.

\par The nucleus can be optically thick to synchrotron self
absorption (SSA) at wavelengths $\sim$ a few mm based on total flux
density measurements \citep{nes95}. At optically thick frequencies,
the flux density is an unreliable surrogate for jet power because it
is never clear if one is seeing a change in SSA opacity or
luminosity. Thus, one wants to sample the nucleus at high frequency
in order to probe optically thin emission that directly relates to
the energy release of the jet. Thus motivated, Figure 6 is a plot of
the $\approx 225$ GHz (1.3 mm) flux density over three decades. The
1.3 mm flux density data after 2002 was obtained at the
Submillimeter Array (SMA) near the summit of Mauna Kea (Hawaii). 3C
84 is included in an ongoing monitoring program at the SMA to
determine the fluxes of compact extragalactic radio sources that can
be used as calibrators at mm wavelengths \citep{gur07}. Observations
of available potential calibrators are from time to time observed
for 3 to 5 minutes, and the measured source signal strength
calibrated against known standards, typically solar system objects
(Titan, Uranus, Neptune, or Callisto). The data was downloaded from
the Sub-millimeter Array Calibrator archive,
http://sma1.sma.hawaii.edu. Before 2002, the data is taken from the
literature \citep{nes95,reu97,tri11}.  Even though 3C 84 has
blazar-like properties, \citet{ver78}, the 1.3 mm light curve has a
predominant background variation on the time scales of many years.
\par We are looking for a diagnostic of jet activity that is contemporaneous with
the H$\beta$ BEL emission. Thus, we want to know the luminosity of
the jet as close to the point of origin as possible. Ostensibly, a
high frequency like 225 GHz rules out optical thin (steep spectrum)
emission and renders the optically thick core the only viable source
of flux. Typically, secondary components are optically thin in the
mm band and can be ignored at high frequency. But this is not the
case during flares in 3C 84. During very luminous flares, the mas
scale structure of 3C 84 is very complex due to the extremely
luminous secondary components (currently C3) that occur on
sub-parsec scales. During the large mm flare in the 1980s, VLBI
observations showed that there was more 89 GHz - 100 GHz flux
density on mas scales than was detected from the nucleus
\citep{bac87,wri88}. There is also evidence during the current flare
that the secondary, C3, is the major contributor to mm flux density
\citep{hod18}. In order to understand the source of the emission
comprising the long term mm wave light curve, we need high
resolution images that can segregate the core from the pc scale jet
emission. No 1.3 mm VLBI observations exist. However, there is a
large history of 43 GHz VLBI observations. The VLBI core flux
densities are limited to observations which have a resolution of
$\leq 0.55\,\rm{mas}\sim 0.6 \,\rm{lt-yr}$
\citep{kri93,dha98,lis01,suz12,nag14,jor17}. We show these data
superimposed on the 1.3 mm light curve in Figure 6 in order to
elucidate the connection.

\par The first 43 GHz VLBI observation that we utilize was based on only 3
stations \citep{kri93}. However, amazingly, these researchers were
able to obtain an image that closely resembles the morphology of the
images from deep 10 station VLBA observations four years later.
Thus, we think that this is a credible result. It has the largest
restoring beam of our sample, a 0.55 mas circular beam. This raises
an issue associated with utilizing a large inhomogeneous sample of
archival data. In spite of all of the observations being 43 GHz with
VLBA (except the \citet{kri93} observation mentioned above) there
are different restoring beams in the dozens of observations that we
acquired \citep{dha98,lis01,suz12,nag14,jor17}. We are looking for
coarse flux variation over time frames of years. Thus, the scatter
induced by the inhomogeneous method of rendering the core flux is
not a deleterious circumstance, yet it is still prudent to process
the core flux densities as uniformly as possible. Most of the
observations have a beam size $\approx 0.2\rm{mas}\times
0.3\rm{mas}$. After 2010.8, the data are obtained from FITS files
that were downloaded from VLBA-BU-BLAZAR Program
\footnote{http://www.bu.edu/blazars}. These data are processed
similarly to the \citet{suz12} in order to maintain a uniform
processing method with our main source of observations. This
eliminates one potential systematic source of scatter in the core
flux density determinations. As with \citet{suz12}, the data is fit
with one circular Gaussian for C1 and one circular Gaussian for C3.
The source has many small components and is very complicated to fit.
As noted in \citet{suz12}, the addition of numerous small components
has a negligible effect on the fits to the major components. Our
Gaussian fits to C1 consistently have a diameter of $0.13- 0.20$ mas
and are brighter than the adjacent jet features. There are
non-negligible ridges of surface brightness just to the south of the
core \citep{nag14}. The implementation of a data reduction technique
that results in a small $<0.2$ mas circular Gaussian for the core
model is an uniform method of segregating the core form these
ridges. We expect this to be a robust method of defining the C1 flux
density. The lone exception is the 1991 observation of \citet{kri93}
in which the circular Gaussian diameter is $\approx 0.8$ mas. Thus,
this data point is likely overestimated relative to the latter
estimates. We also added the core peak intensities that were
downloaded from VLBA-BU-BLAZAR Program
\footnote{http://www.bu.edu/blazars} to Figure 4. This provides an
independent method of interpreting the core luminosity based on the
FITS files. One can see a rise from 2010 toward the 2016-2017 flare
peak and the subsequent decay in both the total core flux density
and the peak intensity.
\begin{figure}
\begin{center}
\includegraphics[width=135 mm, angle= 0]{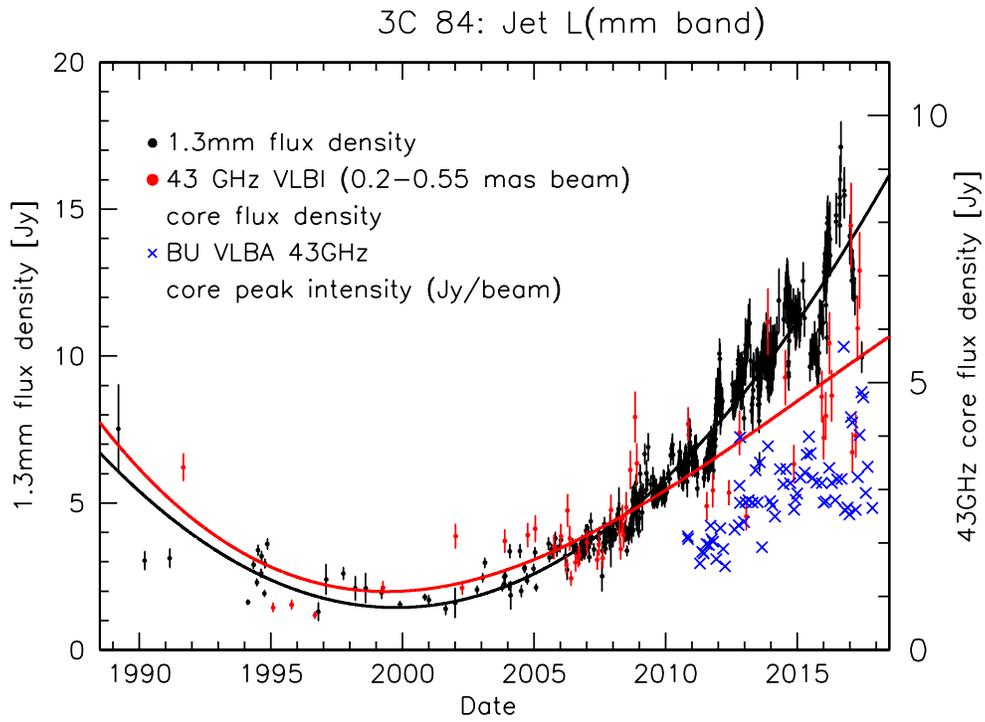}
\caption{A comparison of the 1.3 mm flux density light curve and the
43 GHz VLBI core flux density light curve. The data covers a time
span of $\sim 1990$ until spring 2017. The black curve and the red
curves are the third order polynomial fits to the 1.3 mm and 43 GHz
VLBI data, respectively.}
\end{center}
\end{figure}
\par Figure 6 shows the light curves of both the 43 GHz VLBI core flux density, C(43), and
the $\sim 225$ GHz flux density. The data are selected to cover a
time span of $\sim 1990$ until spring 2017 (a larger data span is
shown in Figures 7 and 8). The red and black lines are the third
order polynomial fits to C(43) and the 1.3 mm flux density,
respectively. The fits help to segregate the long term trends from
the superimposed flickering. It is interesting that there is much
more dispersion about the 43 GHz core trend compared to the 1.3 mm
trend. The two fits are fairly similar from 1991 to 2011. They both
show a broad minimum around 1995 -2002, then an upward trend that is
very close. The upward trend starts to differ significantly in late
2011 when some abrupt flares start to occur at 1.3 mm. From this
time onward, the 1.3 mm light curve shows a significant excess over
what is expected from the 43 GHz VLBI observations of the core.
There is likely a significant contribution from C3 during the height
of flare activity \citep{hod18}. The flux density of C3 started
growing in 2003 and eventually equaled that of the core, C1, in 43
GHz VlBI observations in late 2008 \citep{suz12}. The very high peak
at 1.3 mm in 2015 and 2016 is not indicative of the magnitude of the
core flux density. However, C(43) rises from 2011 to 2017 which does
seem to mirror the 1.3 mm light curve at least qualitatively. It
should be noted that the C(43) light curve does reflect the
2016-2017 peak of the 1.3 mm light curve and the abrupt fading of
the 1.3 mm flux density at the end of 2017, so there is definitely a
strong causal connection. Our 2017 Copernico observation (see Table
1) occurred during this abrupt decay.
\par Based on the high level of 1.3 mm flux density
relative to the total 43 GHz flux density, \citet{nes95}, and
relative to the C1 43 GHz flux density (see Figure 6), there is
evidence that significant SSA absorption exists in the nucleus at
times. Thus, one must always consider this when interpreting the C1
43 GHz flux density. Thus, if the SSA opacity is significant, we
expect a time lag of C(43) relative to changes in the 1.3 mm flux
density. We also note that there could very well be large changes in
SSA opacity at 43 GHz that might also affect the observed C(43) and
this might be the source of the larger scatter from the trend line
at 43 GHz compared to 225 GHz.

\begin{figure}
\begin{center}
\vspace{-2.2cm}
\includegraphics[width=125 mm, angle= 0]{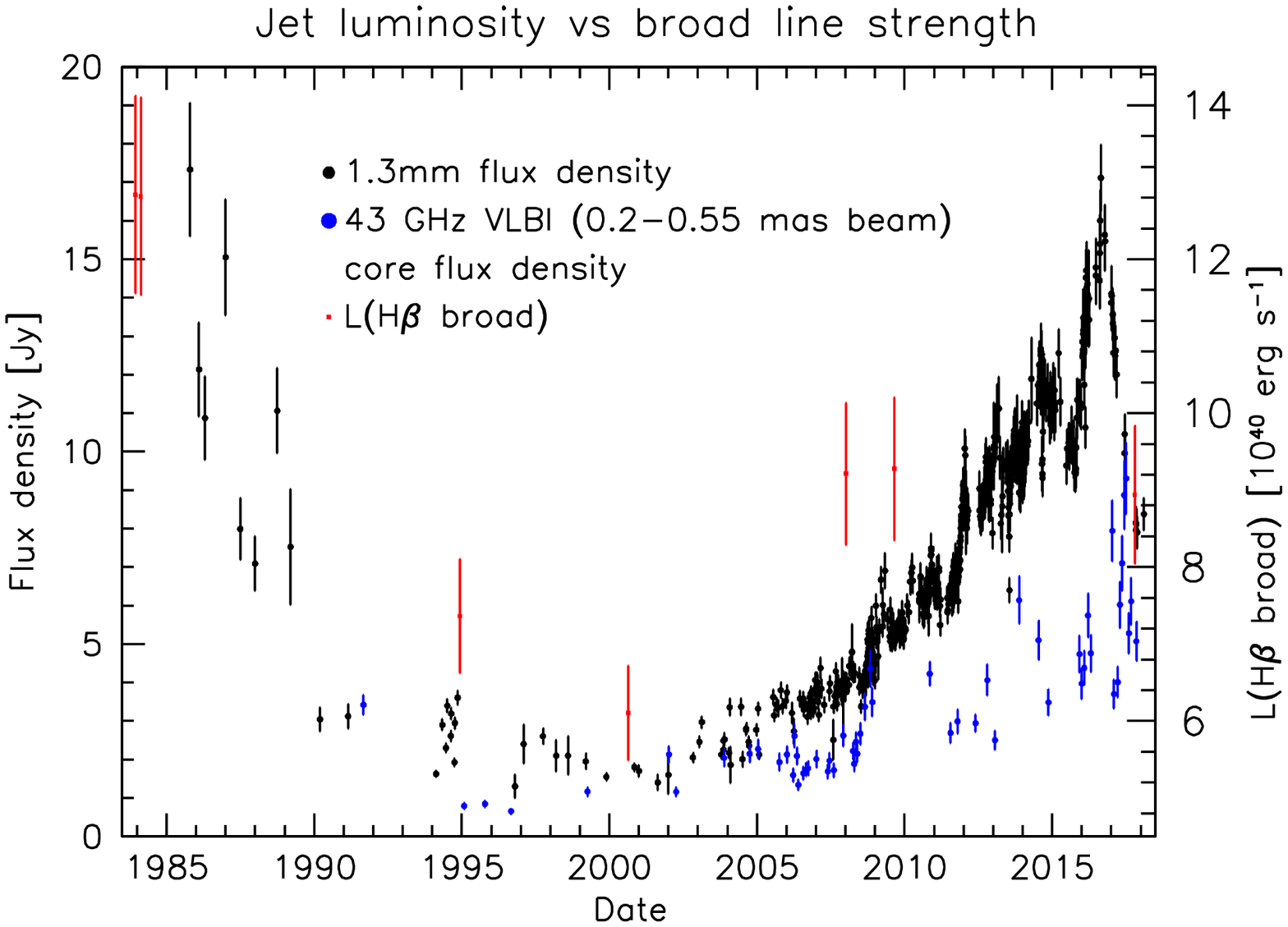}
\vspace{-1.1cm}
\includegraphics[width=125 mm, angle= 0]{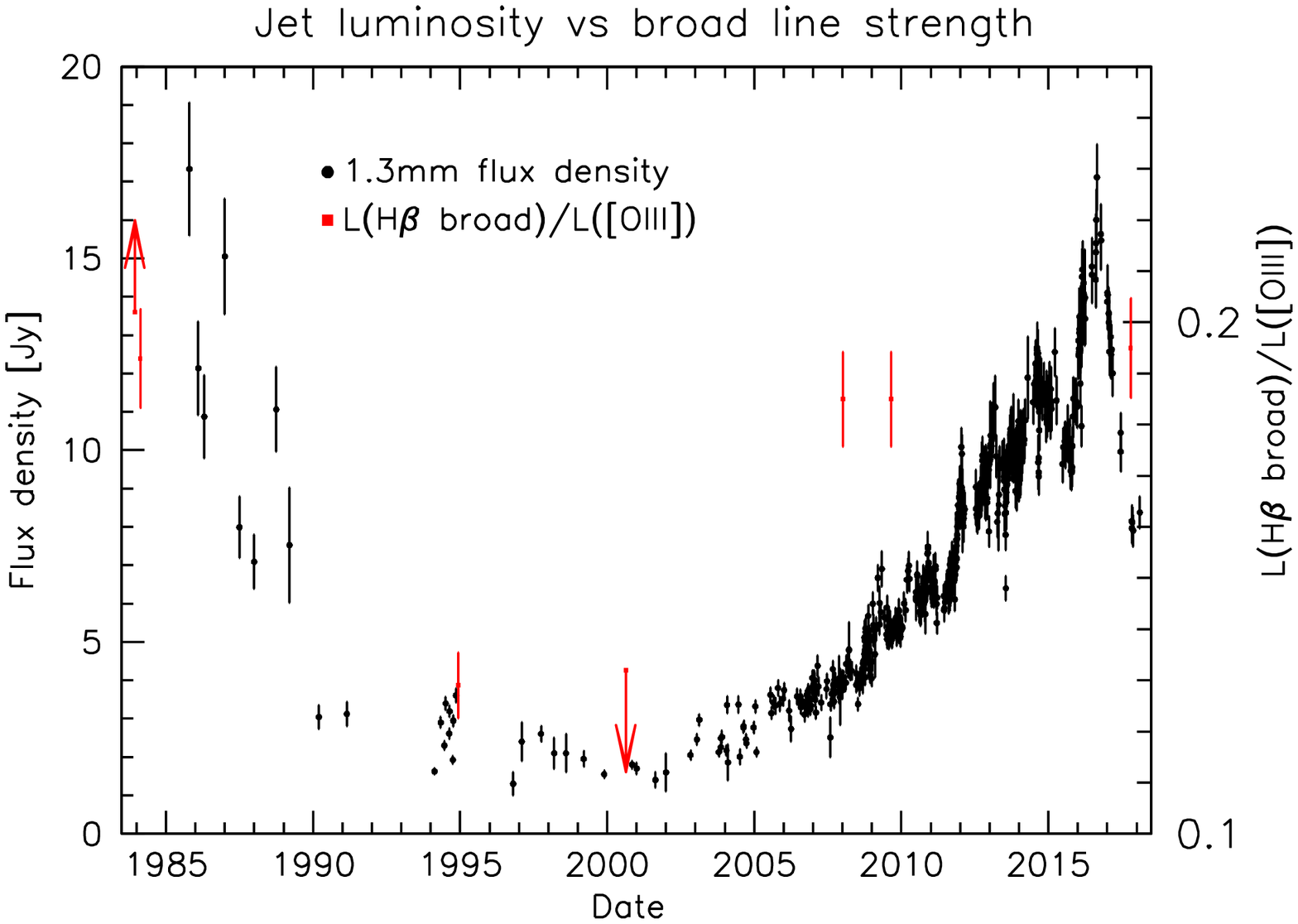}
\caption{\footnotesize{The top frame is a plot of the 225 GHz and 43
GHz VLBI core flux densities versus BEL luminosity from Table 1. The
bottom frame shows the 225 GHz light curve and the L(H$\beta$)
points from the top frame normalized by L([OIII]$\lambda$5007) in a
$6\,(\rm{arcsecond})^{2}$ extraction region.}}
\end{center}
\end{figure}

\par In summary, during states of low to moderate jet activity, the 1.3 mm light curve
agrees with the long term trending of C(43) and is a good surrogate
for the luminosity of the jet on the smallest observable scales
($\sim 0.2$ mas). In states of high flare activity, the 1.3 mm light
curve will over estimate the luminosity of the jet on the smallest
observable scales. The 1.3 mm light curve appears to be a reasonable
surrogate for the core luminosity (but, not necessarily $Q(t)$) from
the late 1980s to 2012.
\subsection{H$\beta$ Broad Line Luminosity and the Luminosity of the
Jet Base} The top frame of Figure 7 appears to show a correlation
between the L(H$\beta$) and the luminosity of the base of the jet.
The changes in L(H$\beta$) are modest, so absolute flux calibration
is critical. One can use the [OII] or [OIII] NL luminosity as an
independent calibration standard, as has been done in reverberation
studies \citep{bar16}. Table 1 shows that the NL luminosity scales
with the size of the extraction region. Understanding the effects of
the size of the extraction and slit PA on emission line fits was the
motivation for our extraction experiments in 2009 and 2017. We need
a robust method of normalizing the size of the extraction region if
we want to use the NL luminosity as a calibration standard.
L([OIII]$\lambda$5007) extracted from a region between $1.69" \times
2" = 3.38\,(\rm{arcsecond})^{2}$ and $2" \times 4" =
8\,(\rm{arcsecond})^{2}$ can be normalized to a $\approx
6\,(\rm{arc\, second})^{2}$ extraction region using our 2009 and
2017 extraction experiments in Table 1 (see the bottom frame of
Figure 7). The HST 2000.663 observation occurs at the important
minimum of jet activity. The large HST [OIII] luminosity relative to
the [OIII] luminosity in the smallest (but much larger) extraction
regions in 2009 and 2017 indicates that centering losses in the
narrow slit were not an issue in the HST observation. The results of
our extraction experiment were designed to normalize the extraction
region with small changes in the extraction size. As such, we cannot
extend this method to the small HST aperture. However, a crude upper
limit in Figure 7 is obtained by ignoring the (significant) [OIII]
luminosity between 0.2" x 0.2" and our smallest extraction regions
of 2" x 2" and 2" x 1.69" \footnote{We average the extraction size
corrections form the combined 2009 and 2017 results. We average the
extrapolation factors from a $3.38\,(\rm{arcsecond})^{2}$ extraction
region to a $6\,(\rm{arcsecond})^{2}$ extraction region and from a
$4\,(\rm{arcsecond})^{2}$ extraction region to a
$6\,(\rm{arcsecond})^{2}$ extraction region. The result is a
correction factor of 1.22. This can be used to form a upper limit on
L(H$\beta$)/L([OIII]$\lambda5007)$ for HST. This is the most
stringent upper limit that we can obtain from our experiment.}.
Similarly, the extraction region in 1983.94 is too large to be
adjusted with our experiment. We create a lower bound by using our
extraction correction from $8\,(\rm{arcsecond})^{2}$ to $\approx
(6\,\rm{arc\, second})^{2}$. The trend in the bottom frame of Figure
7 looks very similar to the trend in the top frame with two
independent methods of flux calibration.

\par If the H$\beta$ BEL gas is photo-ionized by emission from the accretion
flow, as seems likely from the analysis of Section 4, then Figure 7
seems to indicate that the jet luminosity scales with the luminosity
of the ionizing continuum and $L_{\rm{bol}}$. Figure 8 investigates
the possibility that long term trends in $F_{225}(t)$ scale with
$L_{\rm{bol}}$.
\begin{figure}
\begin{center}
\includegraphics[width=135 mm, angle= 0]{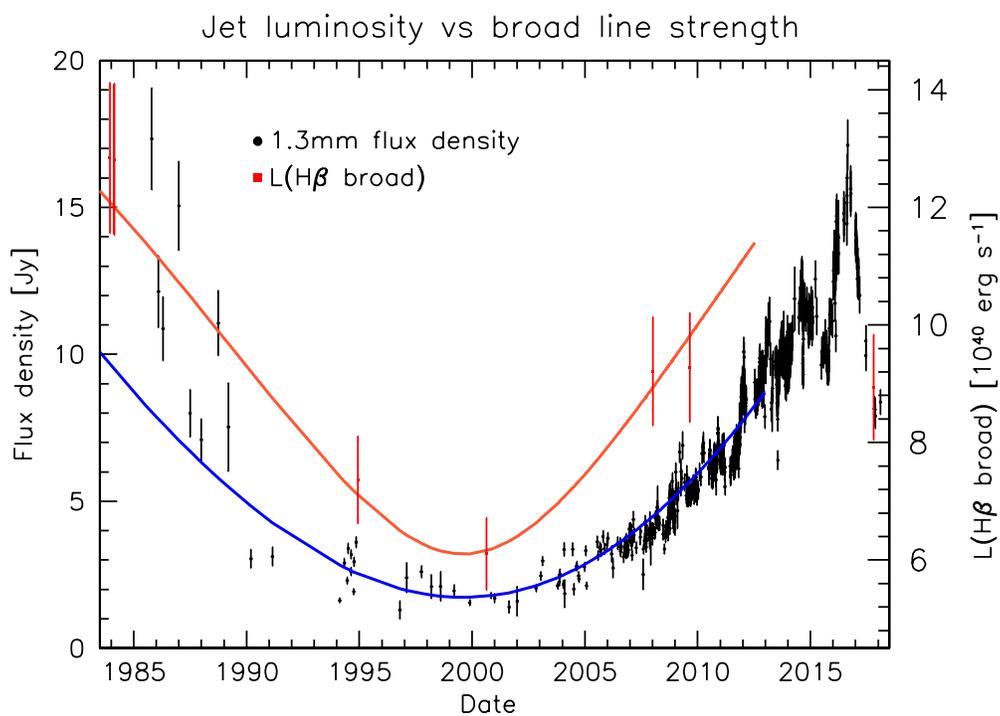}
\caption{\footnotesize{The blue curve is a fourth order polynomial
fit to the 1.3mm light curve, $F_{225}(t)$. Notice that it
underestimates the data before 1988 due the flux from the secondary
component during the strong flare. The orange curve is $L(H\beta)(t)
=\left[\left[\frac{F_{225}(t)}{1.8
\rm{Jy}}\right]^{0.4}\right]6.2\times 10^{40}\rm{ergs/s}$. See the
text for further discussion. Based on Equation (16), $F_{225}(t)$
scales with the strength of the ionizing continuum.}}
\end{center}
\end{figure}
Considering the trends in Figure 6 and the related discussion, it is
concluded that the 1.3 mm flux density is a good representation of
the core luminosity in low and moderate states of jet activity.
Based on 100 GHz VLBI there was still very strong emission in 1987
from the mas scale jet, even though jet activity was decaying
\citep{wri88}. Thus, the earliest point used to fit the long term
trend is 1989.20. The long term fourth order polynomial
approximation to the trend, $F_{225}(t)$, is plotted as a blue curve
in Figure 8. Based on Figure 6, the last point where the 1.3 mm flux
density closely fits the core flux density is chosen to be
approximately 2011.87, just before the small 1.3 mm flare. We
extended the blue curve 6 years to the left although this is very
speculative since it is not being constrained by the data. However,
the extrapolation should be fairly accurate around the time of the
Palomar observations in late 1983 and early 1994 based on 89 GHz
VLBI observations. The core flux density at 89 GHz was 8 Jy in 1983
and 7 Jy in 1984 \citep{bac87}. At 1984.0, our extrapolation of the
blue curve in Figure 8 is 9.5 Jy at 225 GHz. Thus, the extrapolation
gives a reasonable estimate of the core flux density at 225 GHz
considering the 89 GHz flux density and the plausible SSA flat
spectrum of the core at that time.

\par The BEL luminosity varies with the strength of the ionizing
continuum in each photoionized AGN. This is called the responsivity
of the line \citep{kor04}. Empirically, \citet{gil03}, find data to
support the scalings between BEL luminosity and the UV continuum
luminosity, $L_{UV}$,
\begin{equation}
L(H\beta) \propto L_{UV}^{\alpha_H}, \, \alpha_H \approx 0.35
-0.44\;.
\end{equation}
The orange curve in Figure 8, $L(H\beta)(t)$, is motivated by
Equation (16),
\begin{equation}
L(H\beta)(t) =\left[\left[\frac{F_{225}(t)}{1.8
\rm{Jy}}\right]^{0.4}\right]6.2\times 10^{40}\rm{ergs/s}\;,
\end{equation}
where 1.8 Jy is the lowest value of $F_{225}(t)$ in the fourth order
polynomial fit in Figure 8 and 0.4 is the approximate exponent,
$\alpha_{H}$, in Equation (16). The surprisingly good fit combined
with Equation (16) and Equation (17) suggests a dynamic in which
\begin{equation}
F_{225}(t) \propto L_{UV}(t)\;.
\end{equation}
Whether the orange fit is as good as it appears in Figure 8 or not,
the figure strongly supports the notion that the causative agent for
the long term variation in L(H$\beta$) and $F_{225}(t)$ is
$L_{\rm{bol}}$. The good fit could be a manifestation of a
fortuitous cancelation of systematic errors in our method. However,
given the striking nature of the fit, we feel it should be
displayed. The most straightforward physical interpretation of
Equation (18) is that $F_{225}(t)\sim Q(t)$ and $Q(t) \sim
L_{UV}(t)$. We prefer this interpretation as opposed to assuming
that Figures 6 - 8 result from mere coincidence.
\par The blue trend has one sense of curvature, so there is no way
to extend it to the rapid fade in late October 2017 during the
Copernico observation. The 1.3 mm flux density was $\approx 8$ Jy
during the Copernico observation, yet L(H$\beta$) corresponds to a
flux density of about 5 Jy during the rise in 2008-2009. The excess
can be emission form the strong secondary component C3
\citep{hod18}. The excess 225 GHz flux density over what was
expected from the core luminosity was revealed after mid 2011 in the
discussion of Figure 6.
\par In summary, we presented data that indicated a correlation
between L(H$\beta$) and the luminosity of the base of the jet. The
data also supports an interpretation that the connection between the
two quantities is an indirect consequence of the fact that accretion
power is driving the jet and photo-ionizing the BEL. The implication
is that the accretion rate is regulating jet power.
\section{Discussion}
\par The notion that a low accretion rate system can produce
thermal emission from an optically thick accretion flow and
photo-ionize a BEL region has not been anticipated in theoretical
works \citep{nov73,sun89,nar94,yau14}. Thus, it is important to
critique the robustness or our methods and findings. In this regard,
the paper did not disprove other scenarios (besides photo-ionization
of virialized gas) for producing the flux in the broad wings of the
emission lines. In particular, the collisional excitation and
ionization of gas by the jet was not addressed. Utilizing shocks
from a jet collision with gaseous clouds has proved a viable method
of exciting NL gas in some Seyfert galaxies and radio galaxies
\citep{sut93,all08}. However, the BEL gas in NGC 1275 has velocities
exceeding 3000 km/s in the wings and it is not clear if the scenario
can be applied to these high velocities. Before the detailed
calculations of \citet{sut93,all08} were performed, it was
conjectured that the BEL region might be created by a jet
interaction with the surrounding medium in radio loud objects
\citep{nor84}. This idea has not gained traction primarily because
of the development of spectroscopic unification schemes for broad
line AGN. Radio loud broad line objects belong to a class of objects
known as Population B that are defined by H$\beta$ FWHM $\,
> 4000$ km/sec. They tend to have low Eddington rates, $R_{\rm{Edd}}
<0.1$ and include both radio loud and radio quiet objects
\citep{sul00}. Within this population, radio loudness has little or
no effect on the BEL profiles \citep{sul07}. Thus, it has been
generally concluded that the same broad emission line mechanism is
involved in all broad line objects, regardless of whether there
exists a strong jet or no jet. However, one can still question if
this is the case in the very extreme ($R_{\rm{Edd}} \ll 0.1$)
Population B source, NGC 1275. Even though this is not a traditional
explanation of BEL gas in Seyfert galaxies, it cannot be ruled out
due to the very large number of free parameters in the
shock/collisional excitation models that cannot be determined by
observation. Another item to consider is whether the BEL gas is
actually virialized or if it is located within an outflow or inflow.
If there is an explanation other than the photo-ionized/virial
dynamic indicated in Section 4 then there are three puzzling issues
that would need to be explained:
\begin{enumerate}
\item The left hand frame of Figure 3 indicates that $L(H\beta)/\lambda L_{\lambda}(1450 \AA)$ is typical
of what one expects for a low luminosity broad line Seyfert galaxy.
If the gas that produces $L(H\beta)$ is not photo-ionized by
radiation from the accretion source (as is indicated to be true for
the other Seyfert galaxies in the scatter plot based on
reverberation studies) then this circumstance is an unexplained
coincidence.
\item If the gas that produces the BELs is not photo-ionized by
radiation from the accretion source then the fact the CLOUDY
photo-ionization models can simulate the line ratios with physically
reasonable dimension and density is another coincidence.
\item Consider the red excess in the bottom left had panel of
Figure 1 and the BEL fits in Table 1. The wings extend $>3000$
km/sec from the line center. What is the mechanism that drives the
line emitting, ionized gas at $>3000$ km/sec both towards and away
from the observer?
\end{enumerate}
Point 3 addresses the location and origin of the high velocity BEL
gas. If it is virialized gas within the gravitational potential then
points 1) and 2) indicate that it should be photo-ionized by the
accretion flow emission. So there would need to be another physical
location for the BEL gas, hence the invocation of an outflow or
inflow. For outflows, the red excess in the lower left hand panel of
Figure 1 is due to an outflow on the far side of the accretion
plane. Alternatively, the red excess might be due to emission from
an inflow on the near side of the accretion plane. An outflow
component to the virialized gas motion is often used to explain BEL
regions. However, the outflow is driven by radiation pressure and to
achieve gas velocities $>3000$ km/sec requires $R_{\rm{Edd}}> 0.2$,
three orders of magnitude larger than what we found for NGC 1275
\citep{mur95}. In addition, empirically, BEL outflows manifest
themselves as blueshifts in high-ionization lines (e.g., Sulentic et
al. 2007; Richards et al. 2011). Evidence of BEL outflows in Hbeta
is scant, and mainly associated with sources radiating close to the
Eddington limit \citep{neg18}. This is the opposite end of the
Eddington ratio distribution than the Eddington ratio of NGC 1275.
\par Since there is no telescope capable of resolving the BEL gas, one
cannot prove categorically that there is not a jet interaction that
ionizes BEL emitting outflows or inflows on sub-pc to pc scales.
Even so, the scenario is highly challenged because of the
``coincidences" 1) and 2) above that would have no explanation.
Furthermore, the dynamics of the outflow/inflow at $>3000$ km/sec
requires an explanation as well.

\par
The other intriguing finding of our analysis was the apparent
correlation between the luminosity of the H$\beta$ BEL and the mm
band luminosity of the jet base within 0.2 mas of its origin. This
is supportive of a dynamic in which both the jet power and the
intensity of the photo-ionization field scale with the accretion
rate. This is consistent with the magnetically arrested accretion
model of radio jet formation \citep{igu08}. In this model, large
scale poloidal magnetic flux (define $B^{P}$ as the poloidal
magnetic field component) is transported to the inner region of the
accretion flow by the accreting plasma. The inner region of the
accretion flow is in rapid (approximately Keplerian) rotation. The
magnetic flux is dragged around by the rotating plasma, twisting it
azimuthally (creating a toroidal magnetic field $B^{\Phi}$). The
magnetically arrested dynamic also causes the field lines to rotate
with a velocity, $v^{\phi}_{F}$, about the symmetry axis of the
central supermassive black hole in the inner accretion flow. To
external observers, this creates a poloidal electric field
$E_{\perp} =-(v^{\phi}_{F}/c)B^{P}$, equivalently, a cross-field
potential difference. This is tantamount to a poloidal Poynting
flux, $S^{P} = (c/4\pi)B^{\phi}E_{\perp}=-(v^{\phi}_{F}/4\pi)
B^{\phi}B^{P}$, flowing outward from the inner accretion disk,
parallel to the symmetry axis of the black hole, at the expense of
the gravitational potential energy of the plasma. Fully relativistic
magnetohydrodynamic simulations show that the efficiency can be
extremely large (Poynting flux $\sim 0.4\dot{M}c^{2}$, where
$\dot{M}$ is the accretion rate.) if this occurs very near a rapidly
rotating black hole in the active region known as the ergosphere
\citep{pun09}.

\section{Conclusion}
In this article, we described the nature of the weak H$\beta$ BEL in
Section 2 (Figure 1 and Table 1). The situation is made clear by the
coexistence of a P$\alpha$ BEL that also arises from transitions
from the n=4 state with a similar FWHM as shown in Figure 1 and
Table 1.
\par In Section 3, we found that there is a very weak CIV BEL as well.
Section 4.1 showed that the ionizing continuum resembles that of a
weak broad line Seyfert galaxy except that the hard ionizing
continuum is likely somewhat suppressed (Figure 3). This is
consistent with thermal emission from an optically thick accretion
flow not an ADAF even though $R_{\rm{Edd}}\gtrsim 0.0001$. We ran
CLOUDY simulations in Section 4.2 that indicated that the weak, soft
ionizing continuum provides an explanation of the weak CIV BEL
described in Section 3. As a verification of the applicability of
these single zone models of the BEL region and ionizing continuum,
not only is the small BEL luminosity ratio, L(CIV)/L(H$\beta$) $<1$,
replicated, but we also were able to replicate the observed large
L(P$\alpha$)/L(H$\beta$) and L(HeII$\lambda$1640)/L(CIV) ratios.
\par In the process of studying the photo-ionization of the BEL
region we uncovered some other interesting aspects of the accretion
flow. In Sections 4.4-4.6, we found that the virial mass estimates
combined with our CLOUDY simulations indicate that the low
ionization BEL region is being viewed near the polar axis. Combining
this with the optical polarization properties, we concluded that the
innermost jet (within 0.1 pc from the source) is best interpreted as
a slightly off angle BL-Lac jet. In Section 4.7, evidence was also
shown that $L_{\rm{bol}}$ has been fading over decades or centuries.
We point out that NGC 1275 is not the only known LLAGN with a CIV
BEL. NGC 4579 is a LLAGN, with both high ionization and low
ionization BELs \citep{bar01}.
\par The simplified single zone CLOUDY analysis was considered in the context of the BL-Lac
nature of the nucleus and line of sight effects in the the virial
mass estimates. We conclude that the explanation of all of these
results, in consort, favors the following model of the BEL gas: a
density of $\sim 1-3 \times 10^{10} \rm{cm}^{-3}$ located at a
distance of $ \sim 5 \times 10^{16} \rm{cm}$ from the photo-ionizing
source, with approximately solar metallicity.

\par Section 5 presented data showing that $L(H\beta)$ correlates with
the luminosity of the jet less than 3 light months (0.2 mas) from
the central engine. We explored the implications to the dynamics of
jet formation. It appears that accretion power is driving the jet
and photo-ionizing the BEL. This was interpreted in terms of
magnetically dominated accretion models in Section 6.
\par Based on the potential theoretical importance of this
connection between the accretion flow and the jet power, it would be
of significant scientific value to monitor $H\beta$ and $P\alpha$
every two or three weeks. The 225 Ghz luminosity is monitored
regularly and 43 GHz VLBA observations occur monthly. A
``reverberation mapping" of the jet and BELs might provide more
information on the location of the line emitting gas relative to jet
base.
\begin{acknowledgements}
We would like to thank Ski Antonucci and Patrick Ogle for
discussions and insight into the nature of the ionizing continuum.
Kirk Korista also shared his knowledge of photo-ionization of broad
emission line clouds. We thank Barbara Balmaverde for sharing the
details of the data reduction of the Chandra X-ray data. Jeffrey
Hodgson also shared his extensive VLBI knowledge of this source, for
which we are thankful. This study makes use of 43 GHz VLBA data from
the VLBA-BU Blazar Monitoring Program VLBA-BU-BLAZAR funded by NASA
through the Fermi Guest Investigator Program
\footnote{http://www.bu.edu/blazars/VLBAproject.html}. The VLBA is
an instrument of the Long Baseline Observatory. The Long Baseline
Observatory is a facility of the National Science Foundation
operated by Associated Universities, Inc. The results are based in
part on observations collected at Copernico telescope (Asiago,
Italy) of the INAF - Osservatorio Astronomico di Padova. Data was
published with permission of the Submillimeter Array Calibrator
website http://sma1.sma.hawaii.edu/callist/callist.html. The
Submillimeter Array is a joint project between the Smithsonian
Astrophysical Observatory and the Academia Sinica Institute of
Astronomy and Astrophysics and is funded by the Smithsonian
Institution and the Academia Sinica.

\end{acknowledgements}


\begin{thebibliography}{}
\bibitem[Abdo et al.(2010)]{abd10} Abdo, A. A., Ackermann, M., Ajello, M., et al. 2010, ApJ, 720, 912
\bibitem[Allen et al.(2008)]{all08} Allen, M. G., Groves, B. A., Dopita, M. A., Sutherland, R. S.,  Kewley, L. J. 2008, ApJS, 178, 20
\bibitem[Angel and Stockman(1980)]{ang80} Angel, J. and Stockman, H. 1980, ARA\&A 18,
321
\bibitem[Antonucci(1993)]{ant93} Antonucci, R.J. 1993, Annu. Rev. Astron. Astrophys. 31 473
\bibitem[Asmus et al.(2014)]{asm14} Asmus, D., Honig, S., Gandhi, P., Smette, A., Duschl, W. 2014, MNRAS, 439,
1648
\bibitem[Backer et al.(1987)]{bac87}Backer, D.,  Wright, M., Plambeck, R. et al. 1987, ApJ, 322,
74
\bibitem[Barvainis.(1987)]{bar87}Barvainis, R. 1987, ApJ, 320, 537
\bibitem[Balmaverde et al.(2006)]{bal06}Balmaverde B., Capetti A., Grandi P., 2006, A\&A, 451, 35
\bibitem[Balmaverde and Capetti(2014)]{bal14}Balmaverde, B. and Capetti, A.
2014 A\&A 563 119
\bibitem[Barth et al.(2001)]{bar01}Barth, A., Ho, L., Filippenko, A., Rix, H.-W., Sargent, W. 2001 ApJ,
546 205
\bibitem[Barth and Bentz(2016)]{bar16}Barth, A., Bentz, M. 2016 MNRAS Lett, 458,
109
\bibitem[Boettcher et al.(2013)]{boe13} Boettcher, M., Reimer, A., Sweeney, K. and Prakash, A. 2013 ApJ
768 54
\bibitem[Boroson and Green(1992)]{bor92} Boroson, T. A., and Green, R. F. 1992, ApJS, 80, 109
\bibitem[Buttiglione et al.(2009)]{but09}Buttiglione, S., Capetti, A., Celotti, A., et al. 2009, A\& A, 495, 1033
\bibitem[Buttiglione et al.(2010)]{but10}Buttiglione, S., Capetti, A., Celotti, A., et al. 2010, A\& A, 509,
6
\bibitem[Cardelli et al.(1989)]{car89} Cardelli, J., Clayton, G., Mathis, J. 1989 ApJ 345
245
\bibitem[Corbin (1997)]{cor98} Corbin, M. 1997, ApJ 485
517
\bibitem[Chuvaev(1985)]{chu85} Chuvaev, K. 1985, PAZh 111 803
\bibitem[DeCarli et al.(2011)]{dec11}DeCarli, R., Dotti, M., Treves, A. 2011, MNRAS
413 39
\bibitem[Dhawan et al.(1998)]{dha98} Dhawan, V., Kellermann, K. I., Romney, J. D. 1998, ApJL, 498, 111
\bibitem[Donzelli et al.(2007)]{don07}Donzelli, C., Chiaberge, M., Macchetto, F. D.; Madrid, J. P.; Capetti, A.; Marchesini, D. 2007 ApJ
667 780
\bibitem[Dimitrijevi{\'c} et al.(2007)]{dim07}Dimitrijevic, M. S.; Popovic, L. C.; Kovacevic, J.; Dacic, M.; Ilic, D. 2007 MNRAS
6374 118
\bibitem[Evans et al.(2004)]{eva04} Evans, I. and Koratkar 2004, ApJS 150 73
\bibitem[Ferland et al.(2013)]{fer13}Ferland, G., Porter, R., van Hoof, P. et al. 2013 Revista Mexicana de Astronomía y Astrofísica 49 137
\bibitem[Filippenko and Sargent(1985)]{fil85} Filippenko, A., and Sargent, W.  1985, ApJS, 57,
503
\bibitem[Fujita and Nagai(2017)]{fuj17} Fujita, Y. and Nagai, H. 2017, MNRAS, 465,
L94–L98
\bibitem[Gammie et al.(1999)]{gam99} Gammie, C. F., Narayan, R., Blandford, R. 1999, ApJ, 516, 177
\bibitem[Ghisellini et al (2010)]{ghi10} Ghisellini, G., Tavecchio, F., Foschini, L., Ghirlanda, G., Maraschi, L., Celotti, A. 2010
MNRAS 402 497
\bibitem[Giovannini(2018)]{gio18} Giovannini, G.; Savolainen, T.; Orienti, M. et al. 2018, Nature
Astronomy https://doi.org/10.1038/s41550-018-0431-2
\bibitem[Gilbert and Peterson(2003)]{gil03} Gilbert, K., Peterson, B. M. 2003, ApJ, 587, 123
\bibitem[Greene and Ho(2005)]{gre05} Greene, J. E., \& Ho, L. C. 2005, ApJ, 630, 122
\bibitem[Gurwell et al.(2007)]{gur07} Gurwell, M. A., Peck, A. B., Hostler, S. R., Darrah, M. R., \& Katz, C. A.2007 in \emph{From Z-Machines to ALMA: (Sub)Millimeter Spectroscopy of
Galaxies} Astronomical Society of the Paci?c Conference Series, 375,
234.
\bibitem[Hardcastle et al.(2009)]{har09} Hardcastle, M., Evans, D. and Croston, J. 2009, MNRAS, 396,
1929
\bibitem[Ho et al.(1995)]{lho95} Ho, L. C., Filippenko, A., and Sargent, W.  1995, ApJS, 98,
477
\bibitem[Ho et al.(1997)]{lho97} Ho, L. C., Filippenko, A., and Sargent, W.  1997, ApJS, 112, 315
\bibitem[Hodgson et al.(2018)]{hod18} Hodgson, J., Rani, B., Lee, S.-S. et al. 2018, MNRAS 475 368
\bibitem[Humason(1932)]{hum32} Humason, M.  1932, PASP, 44, 381
\bibitem[Igumenshchev(2008)]{igu08}Igumenshchev, I. V. 2008 ApJ 677 317
\bibitem[Jarvis and McLure(2006)]{jar06}Jarvis, M., McLure, R. 2006 MNRAS 369 182
\bibitem [Jorstad et al.(2017)]{jor17}Jorstad, S., Marscher, A., Morozova, D., et al. 2007, ApJ, 846, 98
\bibitem [Kaspi et al.(2005)]{kas05}Kaspi, S., Maoz, D., Netzer, H., et al. 2005, ApJ, 629, 61
\bibitem [Khachikian and Weedman(1969)]{kha74}Khachikian, E., Weedman, D. 1974 ApJ 192 581
\bibitem [Kishimoto et al.(2007)]{kis07}Kishimoto, M., H¨onig, S. F., Beckert, T., Weigelt, G. 2007, A\&A, 476, 713
\bibitem [Korista et al.(1997)]{kor97}Korista, K. T., Baldwin, J., Ferland, G., Verner, D. 1997, ApJS, 108, 401
\bibitem [Korista and Goad (2004)]{kor04}Korista, K., Goad, M. 2004, ApJ, 606, 749
\bibitem [Krichbaum et al.(1993)]{kri93}Krichbaum, T., witzel. A., Graham, D. et al. 1983 A \& A
275 375
\bibitem[Laor et al.(1997)]{lao97} Laor, A., Fiore, F., Elvis, M., Wilkes, B. J.,McDowell, J. C. 1997, ApJ, 477, 93
\bibitem[Lawrence et al.(1996)]{law96} Lawrence, C. et al. 1996, ApJS 107
541
\bibitem[Leipski et al,(2009)]{lei09}Leipski, C., Antonucci, R., Ogle, P., Whysong, D., 2009 ApJ, 701,
891
\bibitem[Lind and Blandford(1985)]{lin85}Lind, K., Blandford, R.1985 ApJ 295 358
\bibitem[Lister(2001)]{lis01} Lister, M. L. 2001, ApJ, 562, 208
541
\bibitem[Mahadevan(1997)]{mah97}Mahadevan, R. 1997, ApJ 477, 585
\bibitem[Marconi and Hunt(2003)]{mar03}Marconi, A., Hunt, L.K. 2003, ApJL, 589, 21
\bibitem[Marziani et al.(1996)]{mar96}Marziani, P., Sulentic, J., Dultzin-Hacyan, D., Calvani, M., Moles, M. 1996, ApJS 104 37
\bibitem[Marziani et al.(2009)]{mar09}Marziani, P., Sulentic, J., Stirpe, G., Zamfir, S., Calvani, M. 2009, A\&A 495
83
\bibitem[Mathews and Ferland(1987)]{mat87}Mathews W. G., Ferland G. J., 1987, ApJ, 323, 456
\bibitem[Norman and Miley(1984)]{nor84}Norman, C., Miley< G. 1984, A \& A,
141, 85
\bibitem[Murray et al.(1995)]{mur95} Murray, N. et al 1995, ApJ 451 498
\bibitem[Nagai et al.(2014)]{nag14}Nagai, H., Haga, T., Giovannini, G., et al. 2014, ApJ, 785, 53
\bibitem[Narayan and Yi(1994)]{nar94}Narayan R., Yi I., 1994, ApJL, 428, 13
\bibitem[Negrete et al.(2018)]{neg18} Negrete, C, Dultzin, D., Marziani P. , Esparza, D. ,
Sulentic, J. et al. 2018 2018arXiv180908310N
\bibitem[Nesterov et al.(1995)]{nes95} Nesterov, N., Lyuty, V., Valtaoja, E. 1995, A \& A
296 638
\bibitem[Novikov and Thorne(1973)]{nov73} Novikov, I. and Thorne, K. 1973, in
\emph{Black Holes: Les Astres Occlus}, eds. C. de Witt and B. de Witt (Gordon and Breach, New York), 344
\bibitem[Punsly and Tingay(2005)]{tin05}Punsly, B., Tingay, S. 2005 ApJL 633 89
\bibitem[Punsly et al.(2009)]{pun09}Punsly, B., Igumenshchev, I. V., Hirose, S. 2009 ApJ 704 1065
\bibitem[Quataert et al.(1999)]{qua99}Quataert, E., Di Matteo, T., Narayan, R., Ho, L.C. 1999 ApJL 525 89
\bibitem[Reuter et al.(1997)]{reu97}Reuter, H.-P., Kramer, C., Sievers, A., et al. 1997, A \& ASS, 122, 271
\bibitem[Reynolds et al.(2013)]{rey13}Reynolds, C., Punsly, B. and O'Dea, C. P. 2013, ApJL, 773,
10
\bibitem[Richards et al.(2011)]{ric11}Richards, G,, Kruczek, N., Gallagher, S., Hall, P., Hewett, P.,  2011, AJ,
141, 167
\bibitem[Riffel et al.(2006)]{rif06}Riffel, R., Rodriguez-Ardila, A., Pastoriza, M. G., 2006, A\&A, 457, 61
\bibitem[Rosenblatt et al.(1994)]{ros94}Rosenblatt, E., Malkan, M., Sargent, W.; Readhead, A. 1994, ApJS
93 73
\bibitem[Ruff et al.(2012)]{ruf12}Ruff, A., Floyd, D., Webster, R.; Korista, K.; Landt,
H. 2012 ApJ 754 18
\bibitem[Scharwachter et al.(2013)]{sch13} Scharwachter, J., McGregor, P. J., Dopita, M. A., Beck, T. L. 2013 MNRAS 429, 2315
\bibitem[Seyfert(1943)]{sey43} Seyfert, C. 1943, ApJ 97 28
\bibitem[Son et al.(2012)]{son12}Son, D., Woo, J.K., Kim, S. 2012, ApJ
757 140
\bibitem[Spinoglio et al.(1995)]{spi95}Spinoglio, L., Malkan, M. A., Rush, B., Carrasco, L., Recillas-Cruz, E.
1995, ApJ, 453, 616
\bibitem[Stern and Laor(2012)]{ste12}Stern. J. and Laor, A. 2012 MNRAS
426 2703
\bibitem[Stevans et al.(2014)]{ste14}Stevans, M., Shull, M., Danforth, C., Tilton, E. 2014 ApJ 794 75
\bibitem[Sulentic et al.(2000)]{sul00}Sulentic, J., Marziani, P., and Dultzin-Hacyan, D. 2000 ARA\& A 38, 521
\bibitem[Sulentic et al.(2007)]{sul07}Sulentic, J., Bachev, R., Marziani,P., Negrete, C. A., Dultzin, D. 2007
ApJ 666 757
\bibitem[Sutherland et al.(1993)]{sut93}Sutherland, R., Bicknell, G., Dopita, M.
1993 ApJ 414 510
\bibitem[Sun and Malkan(1989)]{sun89}Sun, W.-H., and Malkan, M. A 1989, ApJ 346 68
\bibitem[Suzuki et al.(2012)]{suz12}Suzuki, K., Nagai, H., Kino, M., et al. 2012, ApJ, 746, 140
\bibitem[Trippe et al.(2011)]{tri11}Trippe, S., Krips, M., Pietu, V., et al. 2011, A \& A, 533,
97
\bibitem[Veron(1978)]{ver78}Veron, P. 1978, Nature 272 430
\bibitem[Walker et al.(2000)]{wal00}Walker R. C., Dhawan V., Romney J. D., Kellermann K. I., Vermeulen R. C., 2000, ApJ, 530, 233
\bibitem[Willott et al.(1999)]{wil99}Willott, C., Rawlings, S., Blundell, K., Lacy, M. 1999, MNRAS 309 1017
\bibitem[Wills and Brotherton(1995)]{wil95} Wills, B., Brotherton, M 1995, ApJL 448 81
\bibitem[Wills and Browne(1986)]{bro86} Wills, B.J., Browne, I.W.A. 1986 ApJ 302
56
\bibitem[Wright et al.(1988)]{wri88} Wright, M,, Backer, D., carlstrom, J. et al 1988 ApJL
329 61
\bibitem[Wu et al.(2003)]{wuu04} Wu, X.-B., Wang, R., Kong, M. Z., Liu, F. K.,  Han, J. L. 2004,
A\&A, 424, 793
\bibitem[Yuan and Narayan(2014)]{yau14} Yuan F., Narayan R., 2014, ARA \& A, 52, 529
\end{thebibliography}
\end{document}